\documentclass[final,5p,times,twocolumn]{elsarticle}

\usepackage{amssymb}
\usepackage{multirow}
\usepackage{caption}
\usepackage[hyphens]{url}
\usepackage{hyperref}
\usepackage{xcolor}
\usepackage[utf8]{inputenc}

\usepackage[utf8]{inputenc} 
\usepackage{fdsymbol}   
\usepackage{newunicodechar} 
\newunicodechar{⌀}{\ensuremath{\diameter}}

\journal{Journal of Information Security and Applications}

\begin{document}

\begin{frontmatter}




\title{A Novel Classification of Attacks on Blockchain Layers: Vulnerabilities, Attacks, Mitigations, and Research Directions}


\author{Kaustubh Dwivedi\corref{CorrespondingAuthor}}
\ead{f20170615p@alumni.bits-pilani.ac.in}
\author{Ankit Agrawal}
\ead{p20190021@pilani.bits-pilani.ac.in}
\author{Ashutosh Bhatia}
\ead{ashutosh.bhatia@pilani.bits-pilani.ac.in}
\author{Kamlesh Tiwari}
\ead{kamlesh.tiwari@pilani.bits-pilani.ac.in}
\cortext[CorrespondingAuthor]{Corresponding Author}

\address{Dept. of Computer Science and Information Systems, Birla Institute of Technology and Science, Pilani, India, 333031}

\begin{abstract}
The widespread adoption of blockchain technology has amplified the spectrum of potential threats to its integrity and security. The ongoing quest to exploit vulnerabilities emphasizes how critical it is to expand on current research initiatives. Thus, using a methodology based on discrete blockchain layers, our survey study aims to broaden the existing body of knowledge by thoroughly discussing both new and known attack vectors inside the blockchain ecosystem.
This survey proposes a  novel classification of blockchain attacks and an in-depth investigation of blockchain data security. In particular, the paper provides a thorough discussion of the attack techniques and vulnerabilities that are specific to each tier, along with a detailed look at mitigating techniques. We reveal the deep dynamics of these security concerns by closely investigating the fundamental causes of attacks at various blockchain tiers. We clarify mitigation methods for known vulnerabilities and offer new information on recently developed attack vectors. We also discuss the implications of quantum computing in blockchain and the weaknesses in the current technology that can be exploited in the future. Our study advances the field of blockchain security and privacy research while also contributing to our understanding of blockchain vulnerabilities and attacks.
This survey paper is a useful tool for readers who want to learn more about the intricacies of blockchain security. It also invites researchers to help strengthen blockchain privacy and security, paving the way for further developments in this dynamic and ever-evolving field.

\end{abstract}




\end{frontmatter}


\section{Introduction}
\label{intro}
The emergence of blockchain technology stands out as a notable milestone over the past two decades. This innovative field integrates diverse computer technologies such as cryptography, consensus mechanisms, and peer-to-peer networks. Such integration enables the creation of a decentralized and tamper-resistant public ledger, where data and transactions can be stored securely through cryptographic methods without the intermediaries’ need. This characteristic fosters openness, security, and trustworthiness. Initially introduced in 2008 to record Bitcoin transactions, blockchain technology has since facilitated the development of numerous other cryptocurrencies, including Ethereum, Litecoin, Altcoin, and Ripple. Beyond cryptocurrency applications, blockchain has broader utility. It is a distributed database that stores information and records transactions in an immutable chain. A peer-to-peer network upholds this blockchain, ensuring data privacy through cryptographic measures.

Projected to rise from 4.9 billion USD in 2021 to a staggering 67.4 billion USD by 2026, the blockchain market demonstrates remarkable growth potential  \cite{zhang2019security}. Substantial investments in research and development of blockchain technology by financial giants like Goldman Sachs, Morgan Stanley, and Citibank, as well as other prominent financial institutions and internet conglomerates  \cite{zhang2019security} fuel this trajectory. In 2015, Bitcoin garnered the distinction of being the best-performing currency  \cite{its_official_2015_2016}. It was subsequently recognized as the highest-performing asset in 2016  \cite{bitcoin_2016_performance}. Governments worldwide have seized upon the blockchain paradigm, issuing white papers and propelling its evolution. We saw an important moment on September 7th, 2021, when El Salvador adopted Bitcoin as its official legal tender—a pioneering move on the global stage  \cite{elsalvador}.

Beyond its association with cryptocurrencies, blockchain technology enjoys expansive visibility for diverse applications. From secure sharing of medical records  \cite{farouk2020blockchain}, data collection framework  \cite{wang2019survey}, and monitoring supply chains and logistics \cite{sezer2022tppsupply} to revolutionizing virtual circuit devices  \cite{deebak2022robust} and powering NFT marketplaces  \cite{ferone2022blockchain}, blockchain's versatile potential comes to the forefront. A transformative phase, known as the Blockchain 2.0 era, saw smart contracts—a significant leap enabling trustless data exchanges without requiring intermediaries. Ethereum, a prominent Smart Contract-enabled blockchain, stands out. Its market capitalization had soared past 200 billion dollars by mid-2021, underscoring its significance  \cite{ethereum_market_cap}.

Given the expansive size, intricate scale, and diverse applications facilitated by blockchain technology, user apprehension regarding security issues is visible. Notably, several instances of significant attacks on blockchain have already transpired:
\begin{itemize}
\item In March 2014, Bitcoin's transaction mutability vulnerability was exploited to orchestrate an attack on the renowned Mt. Gox exchange platform  \cite{gox_tragedy_2023}. This exploit culminated in the exchange's collapse and the robbery of 450 million US dollars. 
\item In June 2016, malevolent actors directed their attention toward the Decentralized Autonomous Organization (DAO), a smart contract entity. The attackers capitalized on a vulnerability arising from recursion calls within the smart contract, leading to the misappropriation of approximately 60 million dollars.
\item In December 2021, hackers compromised the encryption of two hot wallets linked to BitMart crypto exchange \cite{scharfman2023decentralized}, due to hacked private keys. The hackers were able to steal assets worth 150 million US dollars.
\item In May 2022, Terraform, a decentralized finance service, experienced a cryptocurrency bank run \cite{briola2023anatomy} due to a vulnerability in its protocol, which the hackers took advantage of. Terra, the third largest cryptocurrency ecosystem, soon lost over 50 billion USD in valuation. 
\item Most recently, in November 2022, FTX, the then-third largest cryptocurrency exchange platform, went bankrupt \cite{jalan2023systemic}, leading to a loss of around 9 billion USD. This happened after an article by CoinDesk highlighted the funds' mismanagement issues in the exchange.
\end{itemize}

The potential of blockchain in revolutionizing several industries has led to massive growth. This, however, leads to a significant increase in vulnerabilities that attackers can exploit. Moreover, due to its intricate infrastructure, the knowledge of blockchain and its functioning is mainly limited to academia. This warrants a deep dive into the different vulnerabilities of blockchain, the attacks that aim at these vulnerabilities, and the possible mitigation steps we can take to ensure the robustness of this ecosystem. In this paper, we explore various possible attacks that plague blockchain and classify them based on the layer they target and the specific vulnerability in that layer that they exploit. This classification provides a transparent overview of the shortcomings that need to be tackled in blockchain before its mass adoption occurs. A greater understanding of these vulnerabilities and the different attack vectors will open the doors for blockchain developers and researchers to prioritize their research and work on handling such situations better. Some mitigation techniques mentioned in this paper can be implemented immediately. However, most of them come at some disadvantage, and further research is required to understand such disadvantages and create a feasible solution around them.

Many scholarly articles have surfaced in recent years, delving into the multifaceted realm of blockchain technology. This corpus of work encompasses various aspects, including security concerns and privacy threats. Table \ref{tab:my_label} serves as a comprehensive summary of the collective contributions from the past five years, shedding light on the evolution of insights in this domain. Ghassan Karame et al.  \cite{karame2016security} explore the dangers and vulnerabilities prevalent in digital systems such as Bitcoin and the safety element of blockchain technology, specifically in Bitcoin. In their exploration, they propose methods for mitigating risks. Joseph Bonneau et al.  \cite{bonneau2014anonymity} was the first to describe Bitcoin and similar digital currencies systematically. Mauro Conti et al.  \cite{conti2018survey} examined several loopholes present in Bitcoin and suggested methods to tackle them.

Many survey papers have also illuminated the landscape of blockchain security in recent years. Table \ref{tab:my_label} serves as a focal point, juxtaposing diverse research papers on blockchain security from the last three years. Dasgupta et al.  \cite{dasgupta2019survey} describe issues present in blockchain along with some prevalent attacks and classify them based on their methodological underpinnings. Zhu et al.  \cite{zhu2018research} describe the attacks in detail, providing a more granular analysis. They classify these attacks according to the susceptibility of the targeted blockchain. They also talk about the open problems present in this domain. Zheng et al.  \cite{zheng2018blockchain} describe the blockchain architecture and discuss the security issues and general techniques used. Zhang et al.  \cite{zhang2019security} discuss the architecture, the security techniques, and the open challenges. Huynh et al.  \cite{huynh2019survey} focused more on the different attacks and provided solutions present in the literature to prevent them. 

Mohanta et al.  \cite{mohanta2019blockchain} provide a detailed discussion of the security issues intrinsic to blockchain technology. Li et al.  \cite{li2020survey} combine game theory and blockchain security and talk about rational smart contracts, game theory attacks, and rational mining strategies. Zhang et al. \cite{zhang2023covert} discuss the various covert channels in blockchains that can be leverages to avoid digital eavesdroppers. Hossein et al.  \cite{hossein2021bchealth} provide a survey about IoT networks and the attacks on them from the perspective of blockchain and provide countermeasures for them. It further delves into security paradigms across different contexts. Chaganti et al.  \cite{chaganti2022comprehensive} discuss the Denial-of-Service attacks in the Blockchain Ecosystem and classify them according to their target vulnerabilities. In  \cite{mccorry2017refund}, the authors propose novel attacks on Bitcoin, exploiting the vulnerabilities in the refund policies and authentication of the BIP70 Payment protocol, which Coinbase and BitPay use, two major payment processors that provide trading infrastructure to over 100,000 merchants. To keep blockchain technology updated with the developments in the field of quantum computing, significant research has been done. Kearney et al. Yang et al. \cite{yang2023survey} provide an overview of post-quantum and quantum blockchains, the two current solutions to protect blockchains from developments in quantum computing. \cite{kearney2021vulnerability} measure the risk exposure of the modern cryptocurrencies to quantum attacks and highlight their specific aspects that need updates. They conclude that the use of quantum cryptography is necessary to make blockchain networks quantum-safe. 

We compare the related works with this survey paper on several parameters, such as blockchain architecture, security techniques, discussion on vulnerabilities, attacks, solutions, attack classification, tabular comparison, and Quantum aspect. The comparison between the considered related works and this paper is shown in Table \ref{tab:my_label}. This paper makes the following specific contributions:
\begin{itemize}
    \item This survey paper first introduces a novel classification of the attacks on the layered architecture of blockchain.
    \item We delineate the attacks through a dual-step approach. We first describe the vulnerability of the blockchain layer it targets and then explain what security patches can help fix the weakness. In this thorough approach, this survey paper uncovers and investigates the security limits, vulnerabilities, issues, and obstacles associated with blockchain, along with its safety concerns.
    \item We discuss the advent of quantum computing and its effects on disrupting the current blockchain technology. We present the different aspects of the technology that are under threat due to quantum computing and talk about the ongoing efforts and possible future directions to make it future-proof.
    \item Based on the literature review and classification in the paper, we provide comprehensive future research directions.
\end{itemize}

We lay out the rest of the survey as given. Section II provides a comprehensive exposition of blockchain technology, encompassing an exploration of its fundamental principles, architecture, consensus mechanisms, and foundational tenets. We dissect the blockchain’s intricate ecosystem, providing readers with a nuanced understanding. Section III delves into the intricacies of the layered structure inherent to blockchain. This layered architecture serves as the primary framework upon which the classification framework of this study is founded. We expound on the rationale behind this classification approach, shedding light on its significance in comprehending attacks on blockchain. Section IV presents an in-depth exposition of prominent attacks encountered within the blockchain landscape. We delineate each attack, accompanied by an exploration of existing defense mechanisms documented within the literature. This comprehensive inventory of attacks and corresponding countermeasures offers readers insights into the multifaceted security landscape of blockchain technology. Section V provides future research direction for the community that will help make future developments in blockchain more robust. Section VI provides the conclusion of the research paper, highlighting all the points covered throughout this paper.

\section{Overview of Blockchain}
\label{overview}
Blockchain operates as a decentralized system that records transactions in immutable public ledgers across a network of nodes facilitated by peer-to-peer connections. Each participating node maintains its replicated ledger copy, obviating the need for a centralized authority. The genesis of blockchain's conceptual framework can be traced back to 2008. It found its inaugural application in 2009 when the cryptocurrency Bitcoin harnessed blockchain technology to log and authenticate its transactions. This pioneering leap was attributed to a figure or group identified as Satoshi Nakamoto \cite{nakamoto2008bitcoin}. Bitcoin leverages the blockchain's unalterable ledger to store and validate its transactional activities. This breakthrough was followed by numerous blockchains tailored to diverse, decentralized cryptocurrencies. Ethereum, AltCoin, Binance, Solana, and a spectrum of alternatives adopted blockchain to underpin their transaction verification and storage mechanisms.

\begin{figure*}
\centering
\includegraphics[width=0.82\linewidth, height=0.61\linewidth]{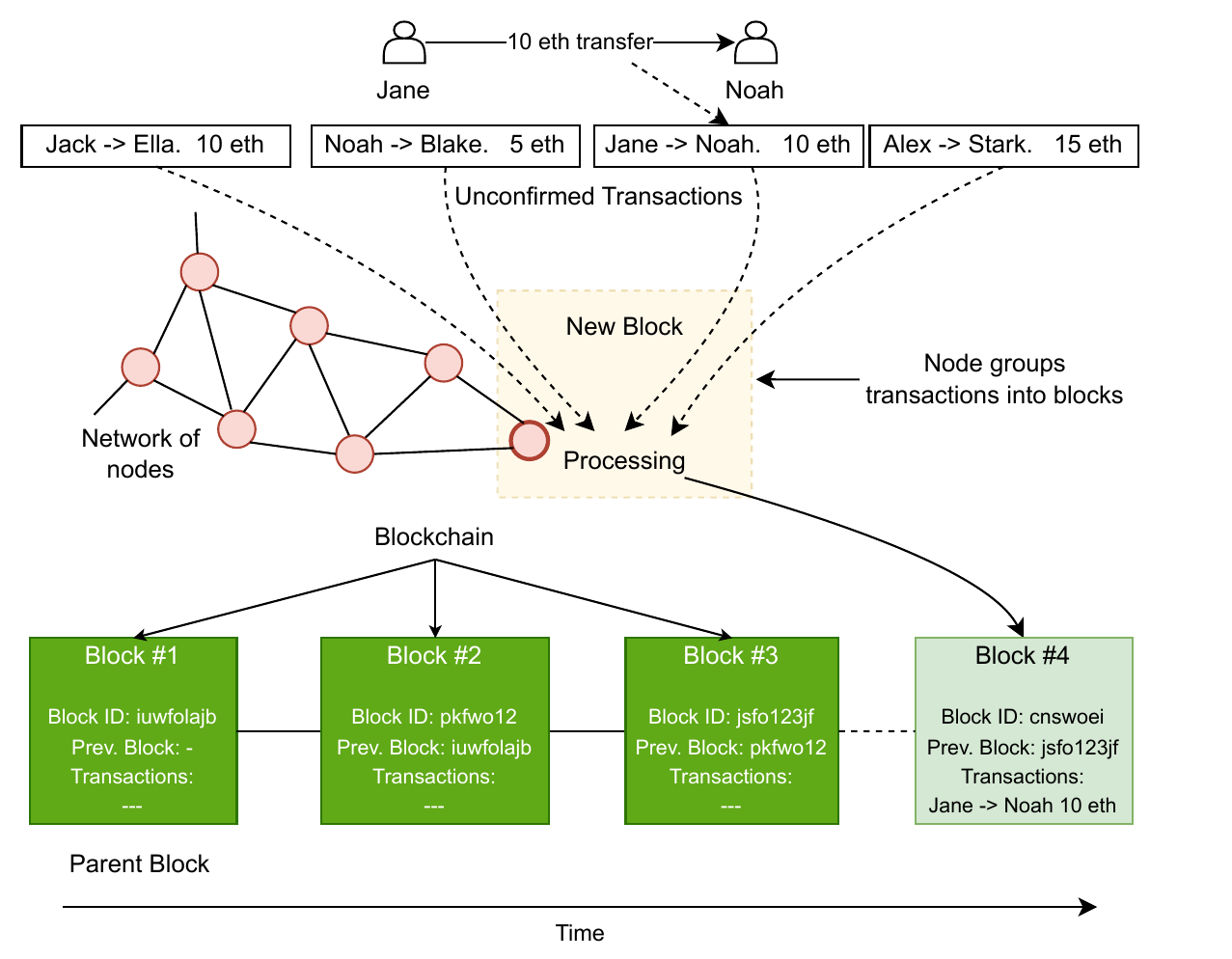}
\caption{Creation of a new block. After a transaction occurs, it gets submitted to the network. This transaction gets to the blockchain within a new block.}
\label{blockcreation}
\end{figure*}

Beyond cryptocurrencies, the transformative potential of blockchain extends into various sectors, such as logistics, education certification, and healthcare. This expansive trajectory paved the way for industries, governments, and academia to display an avid interest in the blockchain ecosystem, leading to its rapid expansion. This proliferation echoes the profound influence of blockchain technology, traversing from its inception in digital currencies to becoming a force driving multifaceted applications across diverse domains.

While blockchain technology might appear straightforward on the surface, its implementation hides a lot of intricacies. Beneath its surface lies a confluence of complexities necessitating an amalgamation of diverse computer science disciplines. The orchestration of this intricate process requires the integration of cryptography, distributed networks, and financial ledger principles. These foundational components coalesce to maintain the seamless functioning of blockchain. The ensuing sections shed light on the inner workings of the technology.

Blockchain materializes as a succession of interconnected blocks, their cohesion facilitated by unique addresses. Each block bears a header housing the preceding block's hash, forging an interlinking sequence. Ethereum introduces variations, incorporating the hash of uncle blocks within the header. Notably, the first block lacks a parent counterpart, consequently omitting a hash within its header. Apart from the previous block hash, the header contains a block-created timestamp, nbits, Merkel tree root hash, block version, and nonce \cite{zheng2017overview}. The block-created timestamp helps in auditing the creation time for each block in the chain. The nonce is a randomly generated number used in hash calculation. All the transactions within the block are kept in a Merkel tree form  \cite{merkley_trees}, and its final hash value is computed and stored in the header as the Merkel tree root hash. Along with the header, every block comprises a body containing the transaction count and the actual transactions.

When a participant within the network initiates a payment transaction, the relevant information becomes integrated into the sequence of blocks housed within a freshly minted block accessible to all network members. This assemblage of transactions. Upon achieving consensus, the blockchain undergoes modification, incorporating the new block into its continuum. Each block in the chain can house hundreds of transactions. Once 51\% of the nodes approve these transactions, the new block can be appended to the existing chain. Noteworthy is the irrevocability of the appended block post-inclusion; its contents remain impervious to alteration. This resolute immutability stems from the network's pervasive possession of the blockchain. In the event of malicious intent, any malevolent action necessitates prevailing over the collective integrity of honest participants. Adding a block to the blockchain heralds the permanence of its encapsulated transactions. 

\begin{figure*}
\centering
\includegraphics[width=1\linewidth, height=0.61\linewidth]{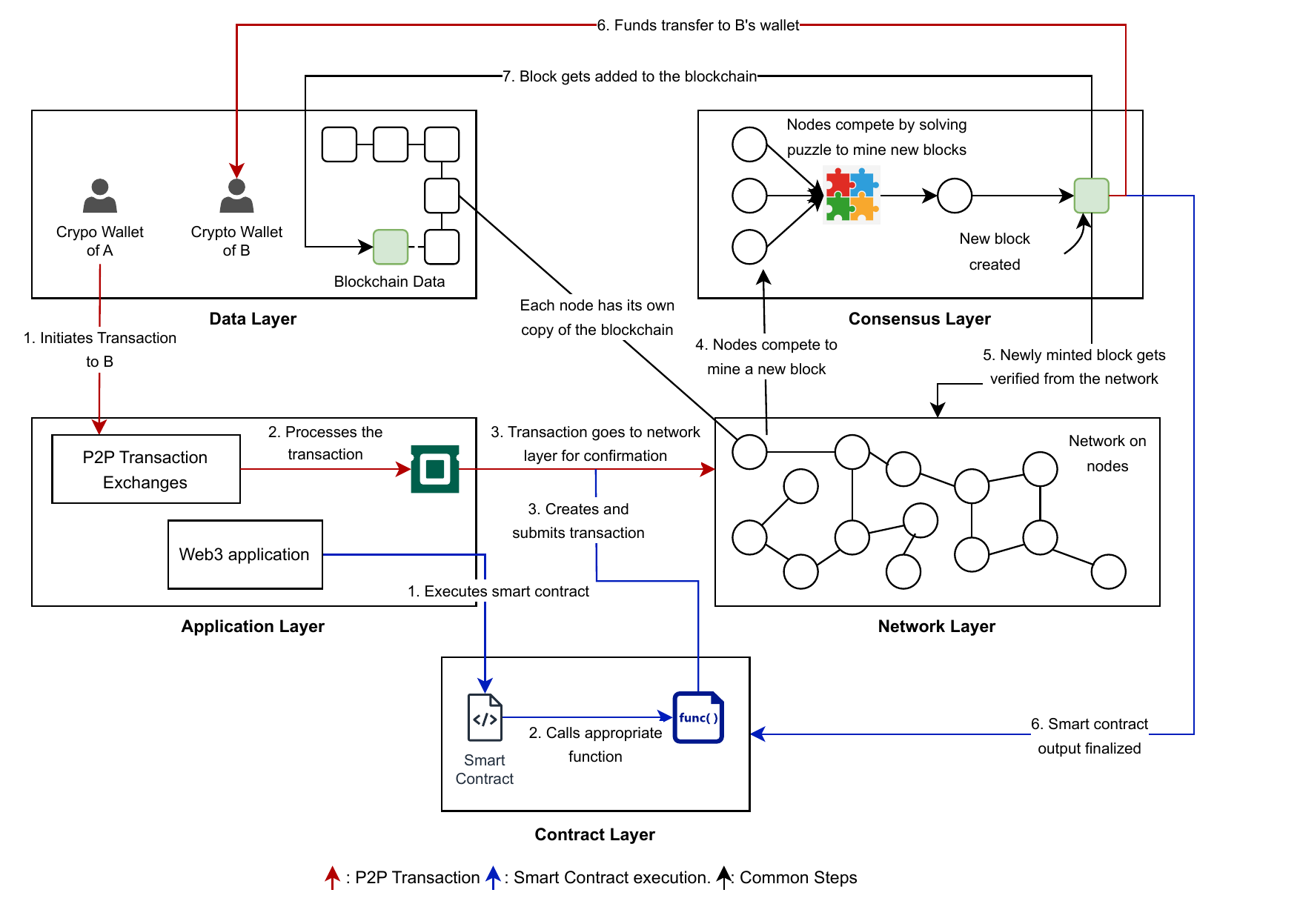}
\caption{Blockchain working based on different layers. Steps are numbered from 1 to 7(Steps 1, 2, 3, and 6 are different for P2P transaction and Smart contracts)}
\label{blockchainlayer}
\end{figure*}

Figure \ref{blockcreation} illustrates the creation process of a block from the execution of the transaction to the permanence achieved through addition to the underlying blockchain. For instance, there is an existing blockchain with block numbers 125 to 128 as the last four blocks, each containing the previous block's hash. When new transactions are executed between various users, they get submitted to the blockchain. The network nodes then work on verifying and grouping these transactions. Once a consensus has been reached by 51\% of the nodes regarding the integrity of the transactions, a new block is created and added to the existing chain as block 129, which will then contain the hash of block 128 along with other details as mentioned above.

As a secure ledger, blockchain intricately arranges these transactions within an expanding sequence of blocks. Upon the culmination of the consensus protocol, new blocks seamlessly integrate into the pre-existing chain, evolving it harmoniously. Beyond the transactional details, each block boasts a unique cryptographic imprint—its hash—which serves as a link to the overarching blockchain. Orchestrated by a distributed consensus mechanism, the network assumes responsibility for many functions: adding freshly minted blocks, verifying transactions before assimilation, and maintaining content uniformity across blockchain copies held by each participant. The inherent strength of a blockchain is manifest in its resolute assurance: once a new transaction secures its place within a block, which is added to the blockchain network, tampering becomes implausible. The inherent immutability of these blocks safeguards this steadfast preservation of transactional integrity once they are integrated. This fundamental characteristic explains how blockchain emerges as a decentralized and unwavering repository, meticulously logging all payments occurring between network users. 

\section{Layered Structure and Attack Classification}
\label{lsac}
Blockchain technology is inherently intricate, demanding substantial time and effort for its development. To comprehensively understand blockchain, dissecting its structure into five distinct layers  \cite{architecture_layer} is beneficial. In this discourse, our focus centers on the architectural classification of attacks predicated upon the specific layer they target and the vulnerabilities within that layer that they exploit. Five fundamental layers underpin the foundation of blockchain's intricate architecture. Figure \ref{blockchainlayer} displays how the blockchain functions based on its various layers.

\subsection{Application Layer} The application layer comprises the 
apps that end users utilize to connect with the blockchain network. The blockchain network acts as the server-side framework for these technologies and interacts with them, serving their needs through API calls. Several attacks are likely on this layer, like Race Attack, Vector76, and Finney Attack, which exploit the fact that some users might accept a 0-confirmed transaction (a transaction that the network has not yet confirmed).

\subsection{Contract Layer} The contract layer encompasses the assortment of contracts created to govern the transactions occurring at the application layer. Given the potential financial implications, we must invest meticulous efforts to ensure the judicious issuance and execution of contracts devoid of potential pitfalls. Prudent execution and integrity are vital prerequisites. Smart contracts, if drafted haphazardly, expose a plethora of vulnerabilities. Cognizant of the potential for mishaps, we must address issues like cyclic calls, erroneous access specifiers, and the acceptance of unauthorized inputs. These frailties pave the way for attacks like Reentrancy Attacks, Short Address Attacks, and more. The contract layer, characterized by its centrality, underscores the importance of diligent contract construction to forestall vulnerabilities and their subsequent exploitation.

\subsection{Consensus Layer} The pivotal layer within every blockchain framework, the consensus layer assumes an indispensable role irrespective of the platform. This layer is responsible for validating, orchestrating the order, and ensuring unanimity concerning the blocks. Its purpose is to safeguard the blockchain network's decentralized ethos, preventing any single entity from dominating the network. At its core, this layer ensures the formation of a harmonious consensus on the truth amongst the participating nodes. The most popular consensus mechanisms available today are proof of work (PoW), proof of stake (PoS), proof of weight, proof of capacity (PoC), proof of authority (PoA), and practical Byzantine fault tolerance (PBFT). Bitcoin uses PoW as its consensus mechanism, where several nodes compete to compute a hash value for the new block. Based on this, the network chooses the node that adds a new transaction as part of the new block in the blockchain. Several vulnerabilities are present within the implementation of this layer. These vulnerabilities, including blockchain centralization and the possibility for forkability, can lead to a number of attacks on this layer. Amongst these are the Shorting Attack, Malicious Reorgs, and the Selfish Mining Attack. Given the important role of the consensus layer in upholding the integrity of the blockchain network, its robustness is significant to ensure that the entire infrastructure is sustained.

\subsection{Network Layer} The network layer, or the peer-to-peer (P2P) layer, is the communication backbone connecting nodes within the blockchain ecosystem. Recognized as the propagation layer, it includes inter-node communication, encompassing discovery, transaction dissemination, and the propagation of blocks. The P2P layer forms the framework through which nodes locate one another, engage in collaborative efforts, and sustain the blockchain's health. Central to this layer's architecture is a P2P network where participants collectively contribute to network operations, leading to shared benefits. At the core of this framework are nodes, categorized into two distinct types: full nodes and light nodes. The former is responsible for transaction verification, validation, mining, and enforcement of consensus rules. Light nodes, in contrast, exclusively retain blockchain headers and facilitate transaction transmission. Yet, several vulnerabilities are present within this layer. The presence of a substantial number of malicious nodes holds the potential to overwhelm the network. This can render target nodes susceptible to isolation and damage their blockchain perspective. These malicious nodes can lead to attacks like the Timejacking Attack, Sybil Attack, and Balance Attack.

\begin{figure*}
\centering
\includegraphics[scale=.63]{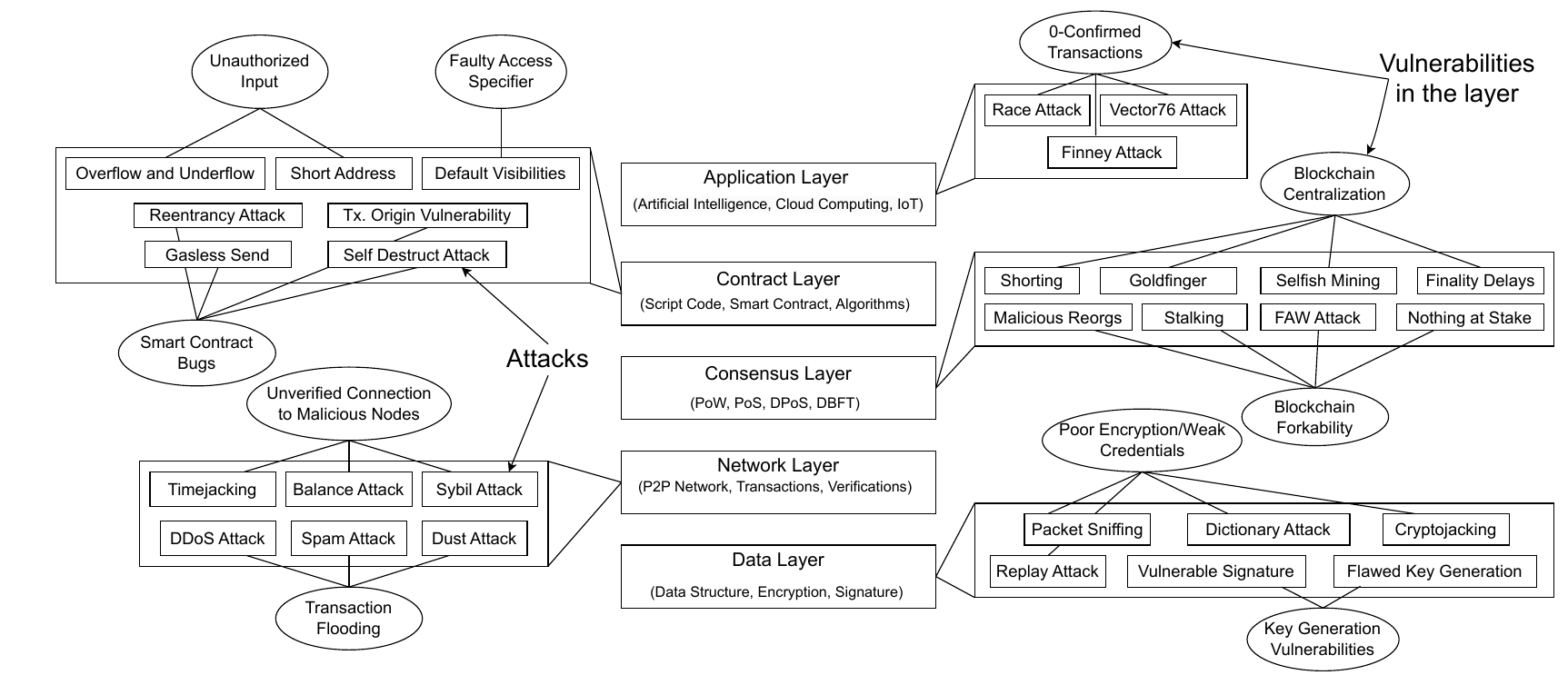}
\caption{Attack classification based on different layers}
\label{attackclassification}
\end{figure*}

\subsection{Data Layer} This layer serves as the repository for data, on-chain, within the blockchain, and off-chain, residing in the database. Safeguarding the security of the data in the blockchain mandates the use of digital signatures for the validation of transactions. Transactions are fortified through digital signatures, which require private keys for payment signing, while their verification is attainable through the corresponding public keys. This mechanism effectively detects any tampering or manipulation of information. Since encrypted data is also signed, any form of manipulation instantaneously invalidates the signature. Encryption further strengthens the data's confidentiality, rendering it invisible to unauthorized users. This combination leads to resilience, fortifying against potential tampering even if the data gets compromised. The application of digital signatures also safeguards the identity of the sender or owner. Notwithstanding these safeguards, the security of the network and users hinges upon the selection of robust, secure credentials. The validity of these credentials becomes important in preventing attacks such as the Dictionary Attack and Replay Attack.

\section{Attacks Description and Mitigation}
This section explores prominent attacks that have surfaced in recent years alongside the advancements made to strengthen the blockchain and mitigate these attacks. We systematically categorize these attacks through a dual-step approach: the initial step involves pinpointing the blockchain layer targeted by the attack, and the next step entails identifying the specific vulnerability within that layer that leads to the attack. We discuss the vulnerabilities, attacks, and their countermeasures in detail. Figure \ref{attackclassification} provides a diagram showcasing the diverse attacks, their corresponding weaknesses, and the layers they target. 

\subsection{Application Layer}

\textit{1)	Zero-Confirmed Transactions}: A zero-confirmation transaction, also known as an unconfirmed transaction, refers to a transaction that has yet to be authenticated on the blockchain. The blockchain's integrity is upheld by the efforts of a decentralized network, collectively engaging in mechanisms to register and validate data residing within this ledger of nodes. When users initiate transactions, they transmit data to the network, which requires validation by one of the network's nodes before incorporation into a block. Since all these blocks are interconnected, validation of each subsequent block inherently ensures the legitimacy of all preceding blocks. A zero-confirmation transaction denotes a transaction not validated by any node \cite{zero_confirmation_coinmarketcap_2023}. Nodes that accept unverified transactions expose themselves to many vulnerabilities, including Race, Finney, and Vector76 attacks. In response to these vulnerabilities, several mitigation steps have been suggested. The most significant is the practice for nodes or vendors delivering services to withhold transactions until they garner validation from participating nodes. This judicious approach ensures that goods or services are extended only once transaction authenticity has been confirmed, reducing the risk of exploitation through attacks.\\

\begin{figure}[t]
\includegraphics[width=1\linewidth, height=0.8\linewidth]{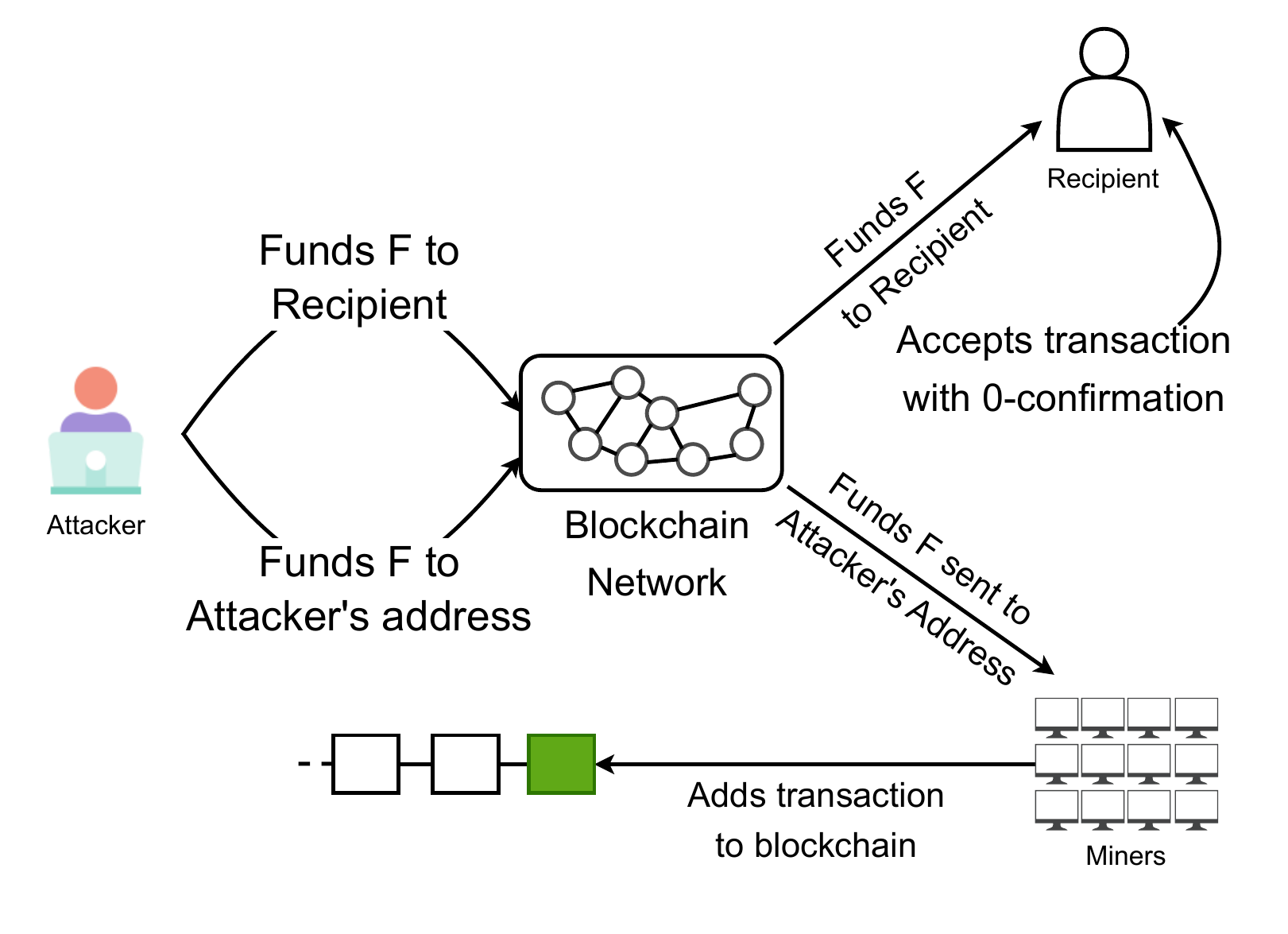}
\caption{Race Attack. The attackers submit two transactions using the same funds, F. The recipient accepts the transaction without waiting for confirmation from the network. Meanwhile, the network accepts the counterfeit transaction.}
\label{raceattack}
\end{figure}

\textbf{\textit{a)	Race Attack \cite{li2020survey}:}} Race attacks refer to a scenario characterized by a contest or "race" between two nearly simultaneous payments or transactions submitted to the network. The objective is to replace an initial transaction, which has yet to be incorporated into the blockchain, with an alternative transaction that channels funds back to a wallet under the attacker's control. This strategy involves the formulation of two distinct transactions: one authentic and the other spurious.
The target of such an attack is nodes that entertain 0-confirmed transactions—transactions that are visible on the network but have not yet been inscribed into the blockchain. The attacker tried to infiltrate the victim node by posing as a legitimate node. Additionally, they may seek a direct or near connection to a mining pool. Subsequently, the attacker dispatches the counterfeit payment to the victim node and a genuine one to the mining pool. The attack succeeds if the target node validates the fabricated transaction and provides goods or services before acknowledging the authentic transaction. Figure \ref{raceattack} provides a diagrammatic depiction of the attack technique.

\textbf{Mitigation}: A good practice is to be patient and wait for a certain number of confirmations before confirming a transaction. This careful approach adds extra protection against possible race attacks and reinforces following standard validation procedures. \cite{architecture_security}.\\

\textbf{\textit{b)	Finney Attack \cite{vokerla2019overview}:}} The Finney attack is the first attack discovered against Bitcoin, credited to Hal Finney \cite{popper2014hal}—an early recipient of Bitcoin payments from the pseudonymous Satoshi Nakamoto. This strategy involves a type of double-spending that requires mining a block beforehand. The attacker engineers an imitation payment concealed within a block and withholds this information from the network. Meanwhile, the coins within this fabricated transaction are deployed to effect a legitimate payment to the target node. The sequence unfolds with the false block being previously mined but unpublicized. Subsequently, upon the delivery of goods or services by the victim node, the attacker does not broadcast the previously mined block. This omission nullifies the transaction executed with the victim node, canceling the resultant payment. Notably, this attack can be executed even if the attacker lacks a majority of the network's hash power (less than 51\%). 

\textbf{Mitigation}: The likelihood of a successful execution reduces proportionately with the attacker's hash power reduction. One effective strategy to mitigate this risk is for the victim node to refrain from validating unconfirmed transactions and wait for verification from established miners. This approach helps prevent the Finney attack \cite{architecture_security}.\\

\textbf{\textit{c)	Vector76 Attack \cite{architecture_security}}:} Another version of double spending attacks, the Vector76 attack targets a vulnerability within Bitcoin's consensus mechanism. In this attack, the attacker exploits the intricacies in the system to steal funds and deceive legitimate users. Through this attack, the malicious actor inserts a double-spent payment into a single block, enabling subsequent exploitation. The attack involves the submission of a self-crafted block to the network, prompting a response confirming its legitimacy. Through this validation, the attacker extracts a predetermined sum before the network identifies the attack. The successful execution of a Vector76 attack hinges upon the attacker's control over two full nodes within the network. One of these nodes establishes an outgoing connection to a service, while the other attempts to forge links with well-connected nodes dispersed throughout the blockchain network. With these connections established, the attacker engineers a block in private and simultaneously crafts two transactions—one of higher monetary value (Transaction A) and another of lesser value (Transaction B). The dishonest miner forwards the block featuring Transaction A to the service connected to one of its nodes alongside the privately created block. Once the service verifies Transaction A, the attacker swiftly withdraws the same sum from the exchange service—mirroring the value of Transaction A. Simultaneously, they dispatch Transaction B to the blockchain network from the second, well-connected node. This sets in motion a chain reaction, leading to a blockchain fork, ultimately annulling Transaction A, the higher-value transaction. Having already extracted funds and nullified Transaction A, the miner emerges with financial gains. The exchange service ends up bearing losses equivalent to Transaction A. 

\textbf{Mitigation}: To effectively counteract these attacks, it is crucial to avoid accepting transactions with only a single confirmation \cite{architecture_security}. Nodes play a vital role in maintaining the network's integrity. They should exercise caution by refusing incoming connections from untrustworthy sources and establishing outbound connections only with nodes that have earned trust. This approach forms a fundamental strategy for mitigating the inherent vulnerability to Vector76 attacks.

\subsection{Contract Layer}

\textit{1)	Unauthorized Input:} Ensuring the security of smart contracts is of paramount significance. When these contracts incorporate inputs from users without strict validation, they inadvertently open doors for many attacks, including Short Address attacks and Overflow and Underflow vulnerabilities. To safeguard blockchain against such attacks, smart contract developers must scrutinize and validate user inputs before deploying them for execution. This proactive safety measure serves as a crucial defense mechanism, effectively preventing potential exploits that may arise from unchecked user inputs.\\

\textbf{\textit{a)	Short Address Attack \cite{short_Address}:}} This vulnerability originates from an inherent imperfection in the Ethereum Virtual Machine (EVM) \cite{short_address_attack_2023}. It results from the EVM's tendency to handle imprecise padding arguments. Adversaries can exploit this susceptibility by providing carefully crafted addresses, creating conditions that can be exploited. Furthermore, a variation of the Short Address Attack appears in the form of a SQL injection flaw \cite{sinha2021detection}. This attack takes advantage of a loophole related to how the EVM responds to underflows. In case of an underflow, the EVM appends a zero to an address to maintain its 256-bit format. Exploiting this, attackers omit the trailing zero from the ether address, introducing a vulnerability in input confirmation. This primarily affects the sender due to the insecure code governing transaction generation. This underscores the complex nature of the Short Address Attack and its intersection with other vulnerabilities, emphasizing the need for robust defensive strategies to prevent its exploitation.

\textbf{Mitigation}: Explicitly specifying the data types and arguments' length in the smart contract will prevent the attacker from submitting unauthorized values. The data provided in the transaction should be checked to match the expected length of the parameters. The use of standard libraries, like OpenZeppelin, in the creation of smart contracts also helps in preventing such attacks through their inherent security checks.\\

\textbf{\textit{b)	Overflow and Underflow:}} This vulnerability targets smart contracts that ingest illicit data inputs  \cite{gao2019easyflow}. Sayeed et al.  \cite{sayeed2019effectiveness} underscore the potential for smart contract overflow when values exceed the permissible threshold. The issue arises when a fixed-size variable tries storing a value beyond the range of its data type. The limitation of contracts, mostly written in Solidity—a language bound by 256-bit value processing—leads to overflow upon the slightest increment beyond this limit. Detecting overflow vulnerabilities within smart contracts goes beyond traditional testing methods. It necessitates the adoption of more advanced approaches to thoroughly examine and proactively address this threat.

\textbf{Mitigation}: A way to mitigate such overflows/underflows is to use standard mathematical libraries to perform operations instead of relying on basic operations.\\

\textit{2)	Smart Contract Bugs:} Without careful authorship, malicious users may exploit many bugs in smart contracts, leading to illicit financial gains or other harmful outputs. Some of the attacks that attack this vulnerability in blockchain that have been uncovered are described below.\\

\begin{figure}
\centering
\includegraphics[scale=0.3]{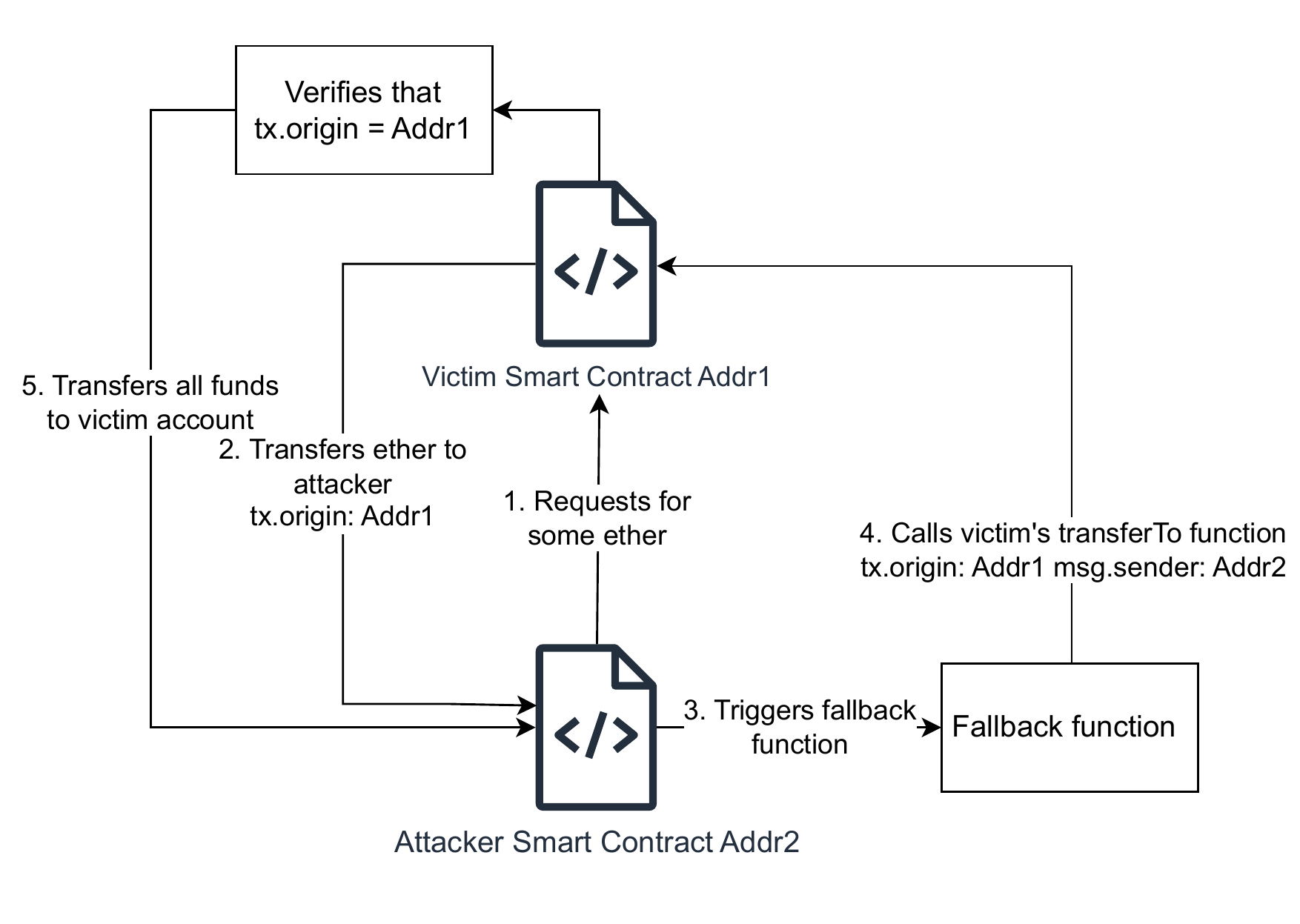}
\caption{Transaction Origin Attack. Steps are numbered from 1(attacker requesting some ether from the victim), to 5(victim losing all their funds to the attacker).}
\label{transactionoriginattack}
\end{figure}

\textbf{\textit{a)	Reentrancy Attack \cite{medium_reentrancy}:}} The Reentrancy Attack is one of the most dangerous threats to smart contracts, capable of causing a complete contract collapse or illicitly accessing sensitive data. This attack typically involves one function making an external call to trigger another contract. This initiation of reentrancy can lead to an attacker invoking a callback, which recursively originates from the attacker's function, creating an unwanted and repetitive loop. Importantly, if a contract includes a "revoke" function, an attacker may exploit it to deplete the contract's balance by repeatedly invoking the function. Attackers can take advantage of two distinct forms of the Reentrancy Attack, using external calls to carry out their malicious actions.

\textbf{Mitigation}: Chinen et al.  \cite{chinen2020ra} propose an innovative solution called the Reentrancy Analyzer (RA), which combines symbolic execution and equivalence verification using a satisfiability modulo theories solver. RA uses Ethereum Virtual Machine bytecodes of smart contracts to explore the interaction dynamics between two contracts, setting itself apart from conventional methods. To prevent this attack, additional measures include proactively updating user balances before initiating transactions to avoid potential loops and carefully labeling untrusted functions to prevent vulnerabilities. This comprehensive approach highlights the seriousness of Reentrancy Attacks and the need for strong defenses to protect smart contracts from their potentially disastrous consequences. Implementing mutex locks to prevent multiple calls to the same function while in an execution state can also help thwart such attacks \cite{rodler2018sereum}.\\

\textbf{\textit{b)	Gasless send \cite{prechtel2019evaluating}:}} The gasless send vulnerability represents an intricacy within the Ethereum smart contracts, originating from the behavior of the "send" function. In this context, when the "send" function transfers Ether to another contract, it triggers the fallback function of the recipient’s contract. One critical aspect of this vulnerability is related to the fixed gas stipend allocated by the Ethereum Virtual Machine (EVM) during this process.

Conventionally, when the "send" function is invoked with a nonzero amount, the gas limit allocated for the recipient contract's fallback function is restricted to 2300 units. This limitation poses a significant challenge when the recipient contract's fallback function involves resource-intensive computations.

The consequence of this vulnerability is seen when an out-of-gas exception is encountered during the execution of the recipient contract's fallback function. Suppose this exception is not effectively checked and propagated within the smart contract. In that case, it opens the door for malicious actors to retain ether, masking their actions behind seemingly normal transactions. It is necessary to address and mitigate the gasless send vulnerability in the Ethereum ecosystem to ensure the integrity and fairness of smart contract interactions. 

\textbf{Mitigation}: Prechtel et al. \cite{prechtel2019evaluating} evaluate this vulnerability in Ethereum smart contracts using the security analysis tool Mythril. They discovered that among the 167,698 smart contracts they analyzed, 57\% were vulnerable to such exploits. Several measures can be undertaken to avoid such exploits. Smart contract developers should perform thorough gas estimation before executing transactions. This involves calculating the gas cost of contract interactions and ensuring that the sender provides sufficient gas to cover the expected execution costs. Tools like Gas Station Network (GSN) \cite{gsn} can help automate gas estimation. Such gas estimations can, however, lead to higher resource wastage for failed transactions[Table 2]. Instead of using the “send” function, using alternate functions like “transfer” or “call.value” allows you to specify the gas explicitly. To handle such cases properly, gas consumption checks should be added to smart contracts. It should, however, be noted that such checks on gas consumption can lead to higher gas fees for executing the transaction.\\

\textbf{\textit{c) Transaction Origin Attack \cite{zhang2020txspector}:}} This attack represents a phishing tactic that could potentially deplete the assets of a contract. In Solidity, tx.origin serves as a global variable that provides the address of the initiating transaction sender. Contracts employing tx.origin for user authorization are susceptible to phishing attacks. Attackers can exploit the utilization of "tx.origin" within authorization logic by deploying a malevolent contract and invoking functions in another contract reliant on "tx.origin". Through this, they can assume the identity of an externally owned account (EOA) and circumvent authorization checks, potentially attaining unauthorized access to the contract's functions and data.

The attacker deploys a smart contract featuring a fallback function that triggers upon receiving Ether. This fallback function then invokes the transfer function of the victim contract, which initially verifies whether "tx.origin" matches the owner of the victim contract. Subsequently, it transfers the specified amount to the designated recipient, often the attacker's address. The fallback function is executed if the attacker persuades the victim to transfer any amount to their contract. Since the transaction's origin traces back to the victim's account, the validation in the victim's function will succeed, thereby deceiving the victim into sending the amount specified in the attack function. The attack is explained graphically in Figure \ref{transactionoriginattack}.

\textbf{Mitigation}: To mitigate such vulnerabilities, it is strongly advised to refrain from using "tx.origin" for authentication purposes, as it merely indicates the initiator of the transaction (which, in our example, was the victim contract itself). Instead, using msg.sender  \cite{tx_origin}, which references the immediate caller of the function (in this case, the attacker's account), ensures proper transaction authorization. However, as mentioned in Table 2, using msg. sender improperly can make the smart contract vulnerable to Reentrancy attacks that have been discussed before since msg.sender hides the actual transaction origin address from the smart contract. Furthermore, within the victim contract's function, fund transfers should not be conducted via the address.call.value(amount)() function. Instead, address.transfer() should be used, as it provides a gas stipend of 2300. This restriction in gas allocation hinders potential attacking contracts from performing further computations beyond emitting events.\\

\textbf{\textit{d)	Self-Destruct Attack \cite{chiu2021mind}:}} Self-destruct is a built-in function in Solidity that effectively removes a contract from the blockchain and sends its remaining ether to a designated recipient. Therefore, when a contract is destroyed, storage space is freed up in the blockchain as its code and data are removed. The self-destruct function is called using the address of the ether recipient as an argument. The recipient’s address will receive all funds held by the contract at the moment of destruction. However, the caller will still have to pay for the gas used to invoke the contract’s self-destruct call. After a contract is self-destructed, all references to it will now point to a bytecode of 0x, just as if it were a regular account. Since the blockchain is immutable, all past transactions and contract calls will remain in the history of previous blocks and cannot be removed even if the contract is destroyed. The contract code is still kept in previous blocks. No assets other than ether (such as tokens) will be sent to the recipient’s address at the moment of destruction, so these will be lost. Any funds and assets sent to the address of a destroyed contract will be lost.

A malicious contract can use self-destruct to force sending ether to any other contract. When a contract is self-destructed, any remaining ether in the contract's balance is sent to a specified address. If an attacker controls the address or is not properly secured, the ether could be redirected to an unintended recipient, resulting in financial loss. In addition to the financial loss, if the conditions for self-destruction are not properly defined or are inadequately secured, an attacker could trigger the self-destruct function to terminate a contract prematurely, resulting in the loss of contract functionality and any assets stored within it. Self-destructing one contract can have implications for other contracts that depend on it. If contracts rely on data or functionality provided by the self-destructed contract, they may become dysfunctional or vulnerable.

\textbf{Mitigation}: To mitigate the self-destruct vulnerability, smart contract developers should carefully define and test the conditions for contract self-destruction to prevent unintended termination. Ensure that any ether transferred during self-destruction is sent to a trusted and secure address. The potential impact on other contracts that rely on the self-destructed contract should also be considered. The condition for self-destruct should have no dependence on the balance in the contract since it can be manipulated by the attacker artificially  \cite{self_destruct}. This function is expected to be deprecated in the upcoming versions of Solidity  \cite{solidity_release}.\\

\textit{3)	Faulty Access Specifier:} Within the domain of Solidity functions, access specifiers, also known as visibility specifiers, control how the invocation mechanism works. \cite{default_visibility}. They also regulate the interaction when external contracts invoke functions through derived contracts. Different programming languages support different numbers of access specifiers. Solidity, for example, currently supports public, internal, external, and private accesses. The "public" specifier allows any external entity to access the function or state variable, whether it's another contract or an external user. The "internal" specifier restricts access to the current contract and contracts derived from it. The "external" specifier is designed explicitly for functions that can only be called externally. Finally, the "private" specifier provides the highest level of access control. Functions or state variables marked as private can only be accessed within the contract they are defined in and are not visible to external contracts or users. Inaccurate utilization of visibility specifiers can yield severe repercussions within smart contracts.

\textbf{\textit{a)	 Default Visibility Attack:}} An essential aspect of the visibility specifier is its influence on the accessibility of external functions, especially when the specifier is inadvertently absent. If not specified otherwise, the visibility defaults to "public." Oversight of specifying the visibility as private can open a window of vulnerability, whereby external contracts can exploit this exposure. In this attack, the attacker targets such smart contracts and tries invoking its functions or data members, which should be hidden from the public. In July 2017, there was an attack on an Ethereum wallet smart contract \cite{multi_sig_wallet} whose initialization function was missing the proper access specifier, defaulting it to public, enabling any attacker to call this initialize wallet function to set the ownership of the wallet to their wallet address and steal all the funds from the victim.

\textbf{Mitigation}: The graveness of this attack necessitates meticulous visibility specifier handling to mitigate the inadvertent susceptibility stemming from default visibility settings. Modifying the default access specifier to private instead of public from solidity can also help reduce the occurrence of such attacks.\\

\subsection{Consensus layer:}

\begin{figure}
\centering
\includegraphics[width=1\linewidth, height=0.7\linewidth]{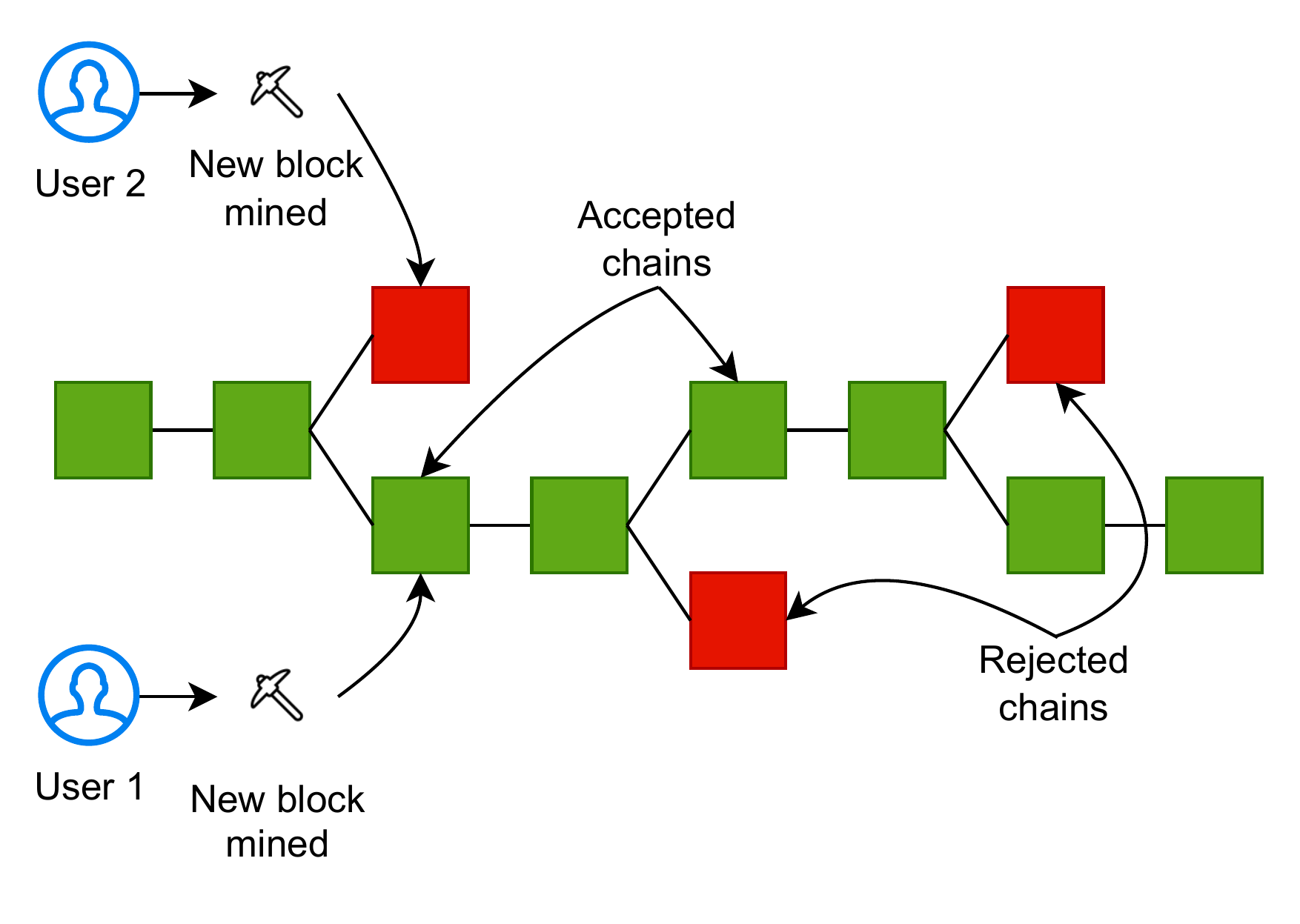}
\caption{Blockchain forking. Mined blocks from User 1 and 2 get accepted by the blockchain simultaneously, leading to a fork. Subsequently, the longer chain is accepted, and fork is resolved}
\label{blockchainforking}
\end{figure}

\textit{1)	Blockchain Forkability:} Nodes in different blockchain systems are inherently designed to synchronize with the longest chain within the current network. This principle leads to forks—instances when multiple miners discover new blocks at nearly identical timestamps. Forks are subsequently resolved as freshly mined blocks are added to the chain, causing one chain to exceed the length of the other, as shown in Figure \ref{blockchainforking}. However, this characteristic leads to vulnerabilities like Malicious Reorgs, FAW attacks, and Stalking Attacks. 

\textbf{\textit{a)	Stalking Attack:}} Chicarino et al.  \cite{chicarino2020detection} introduce the stalker attack, a distinctive variant of the Selfish Mining Attack. Despite both attacks occurring within the same domain, their purposes are significantly different. While the selfish mining attacker seeks illicit financial gains over legitimate miners, the stalker attacker aims to curtail the ability of a node to add a block to the blockchain or insert transactions onto a block set. During a stalking attack, the attacker observes and manages two branches of the blockchain, one reflecting the legitimate chain and the other the private branch of the attacker. Depending on the relative lengths of these branches, the attacker executes one of three steps following the mining or addition of new blocks by other miners.

•	Wait: When the private branch of the attacker surpasses the legitimate branch's length, and the victim node has not added any blocks, the malicious miner continues to work on their private branch.

•	Publish: If the branch of the attacker surpasses the legitimate branch and the victim node uploads a block, the attacker discloses their private branch, nullifying the legitimate one.

•	Adopt: The malicious miner suspends the attack and resumes work on the legitimate chain, determining that the legitimate branch will likely become the longer chain.

Since the stalking attack targets a specific victim node, the lengths of potential forks can diverge even more, resulting in increased fork lengths. As long as the malicious miner detects packets from the victim node in the blockchain, they persist in executing this attack to impede the victim's blocks from being added.

A notable trend is observed in the stalking attack: mean fork lengths tend to rise with an increase in the malicious miner's hash power but decrease with a rise in the victim node's hash power. This shift in fork length demonstrates the nuanced dynamics of the attack. In contrast to the selfish mining attack, the stalker attack showcases a higher frequency of forks due to its distinct objectives.

\textbf{Mitigation}: To prevent the Stalking Attack, two prominent preventive measures have been suggested \cite{selfish_mining_2023}. The first suggests random allocation of miners to divergent branches emerging post-fork, thus ensuring a distributed approach to block extension. The second method involves imposing a threshold on mining pools to impede any miner from gaining an undue advantage over their counterparts in the network. Limiting the concentration of power within a pool can mitigate the potential for malicious exploitation.\\

\textbf{\textit{b)	Malicious Reorgs:}} Neuder et al.  \cite{neuder2021low} delve into two important attacks on the Ethereum 2.0 chain: Malicious Reorgs and Finality Delay. Reorgs, stemming from a fork-choice rule, take place when one branch dominates a prior one, which removes blocks from the latter. These reorgs can occur naturally due to network latency. On the other hand, malicious reorgs are strategic attempts by attackers to exploit them for double-spending or front-running substantial transactions.

Malicious Reorgs are a potent tool for double-spending attacks. Preventing double-spending is critical for the security of cryptosystems. The Bitcoin paper by Nakamoto \cite{nakamoto2008bitcoin} provided a comprehensive analysis of the feasibility of such attacks. However, the likelihood of malicious reorgs diminishes as the length of the reorg increases, implying that transactions deep into the blockchain have a higher probability of remaining intact.

Another consequence of Malicious Reorgs is front-running, which exploits large transactions to identify arbitrage opportunities. This practice is associated with decentralized exchanges in Ethereum. Attackers pay elevated gas fees to ensure their transactions take precedence in block inclusion, allowing them to front-run larger transactions. Malicious reorgs can be used to control the transaction orders, enabling attackers to manipulate the sequence of transactions. As Ethereum 2.0 evolves, addressing these vulnerabilities is of prime importance for maintaining the security and trustworthiness of the platform.\\

\textbf{Mitigation}: The mitigation steps in the Stalking Attack, described previously can be adopted to prevent Malicious Reorgs as well. 

\textbf{\textit{c)	FAW Attack:}} Kwon et al.  \cite{kwon2017selfish} introduce a novel blockchain attack known as the Fork After Withdrawal (FAW) Attack. This attack combines the elements of a Block Withholding Attack with a deliberate fork in a Proof of Work (PoW) blockchain setting. In PoW blockchains, as mining difficulty increases, miners often form mining pools to collaboratively share profits amidst the challenging environment. These pools consist of a pool manager and miner workers. The manager distributes unsolved problems to the miners, who then return Full Proof of Work (FPoW) and Partial Proof of Work (PPoW) shares. Once a block is generated, the manager broadcasts it to the network and allocates profits to the participants.

The FAW attack refers to a malicious actor infiltrating mining pools to exploit their structure. When an external node, not part of the pool, successfully creates a block with transactions, the attacker submits an FPoW to the pool's manager. A fork occurs if the manager accepts the FPoW and propagates a block to the network. If the attacker's block is chosen, the mining pool receives the block reward, which benefits both the pool and the attacker. For a single attacked pool, the attacker consistently gains profit from this attack scenario. 

To maximize gains, the malicious actor can target multiple pools concurrently. If multiple pools participate in the attack, the investigation reveals that the attacker still reaps benefits, although it varies in magnitude based on pool parameters. In cases where two mining pools engage in mutual attacks, the pool with greater mining power tends to achieve higher benefits. Consequently, a Pareto optimum equilibrium \cite{pardalos2008pareto} emerges in the FAW attack game when multiple pools strategically launch the attack.

\textbf{Mitigation}: The mitigation steps in Stalking Attack, described previously can be adopted to prevent FAW Attacks as well.

\begin{figure}
\centering
\includegraphics[scale={.6}]{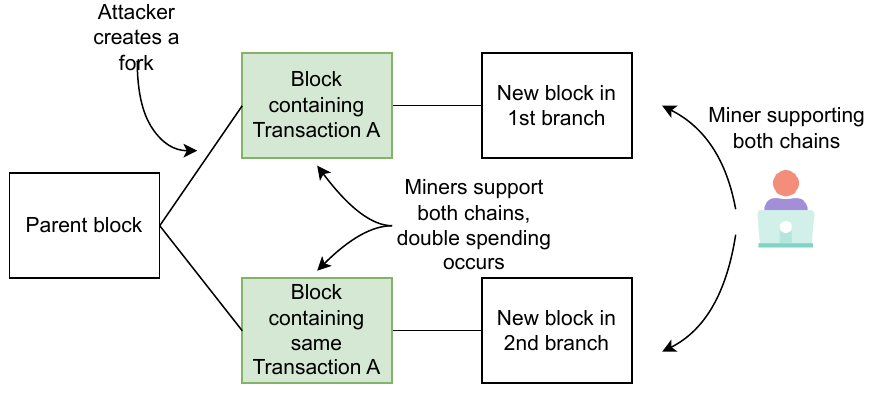}
\caption{Nothing-At-Stake Attack. The attacker forks a blockchain. Since miners are supporting both chains, it leads to double spending opportunities.}
\label{nothingatstakeattack}
\end{figure}

\textbf{\textit{d)	Nothing-At-Stake \cite{nothing_at_stake_theory_2023}:}} The rise of Proof-of-Stake blockchain networks has introduced a new set of consensus mechanisms and security considerations. The “Nothing-at-Stake” attack has emerged as a critical concern. In PoS blockchains, validators, also known as stakers, are responsible for block validation and creation based on the amount of cryptocurrency they “stake” as collateral. Unlike Prood-of-Work systems, where miners invest significant resources in hardware and electricity, PoS validators face a unique dilemma during blockchain forks. Figure \ref{nothingatstakeattack} illustrates the "Nothing-At-Stake" attack.

Whenever the network splits into multiple branches due to consensus disagreement or protocol upgrade, PoS validators must decide which branch of the fork to support. Unlike PoW miners, they do not invest in costly hardware or electricity and hence support multiple branches simultaneously. It is in the best interest of all the validators to continue working on both the chains of the forks. If one validator works on only one chain, and the other one becomes longer, they will not extract any profit from any work that they did on the shorter chain since it would be discarded. A malicious actor might exploit this dilemma by supporting multiple branches, engaging in double spending attacks by creating conflicting transactions on different branches, and effectively spending the same cryptocurrency twice. After forking the blockchain, they would mine only on their fork, and other actors would mine on both forks. The attacker’s fork likely becomes the longest chain. This would not require the attacker to own a huge stake in the system. Supporting multiple branches can lead to network confusion, transaction reversibility, and network instability, potentially compromising security and reliability.

\textbf{Mitigation}: There have been no major Nothing-At-Stake attacks in the world; however, big PoS blockchains like Ethereum are trying to address this vulnerability in Casper \cite{prashantmigration}, their PoS protocol  \cite{nothing_at_stake_theory_2023}. Casper requires the validators to submit a kind of security deposit, which will be the basis of the consensus protocol. This disincentivizes mining on various chains of the forks, thus removing the vulnerability as it introduces a big cost for a dishonest validator. Such security deposits, however, might become a barrier to entry for newer participants in the blockchain. Many PoS blockchains implement slashing mechanisms; penalizing validators caught supporting multiple branches or engaging in malicious behavior. Slashing involves confiscating a portion of the staked cryptocurrency of the validator as a penalty. PoS protocols can also incorporate mechanisms to achieve rapid transaction finality, reducing the likelihood of forks and related vulnerabilities.\\ 

\textit{2)	Blockchain Centralization:} One crucial aspect highlighting the significance of blockchain technology is its inherent decentralization, enabling trustless transactions and record-keeping without the need for a central authority. However, a concerning situation arises when a blockchain network encounters a concentration of hashing power among a subset of miners. This phenomenon, termed blockchain centralization, can engender many vulnerabilities, including Shorting Attack, Selfish Mining, and Finality Delays. These exploits are rooted in the ability of concentrated entities to exert disproportionate control over the network's operations. 

\textbf{\textit{a)	Shorting Attack \cite{lee2020short}:}} Broadly characterized, a Shorting Attack entails strategically selling an asset with the anticipation of its imminent value decline. The seller lends the asset at the current market rate, committing to repurchase the same asset in an equivalent quantity from the buyer. The crux of the scheme lies in the price disparity between the initial sale and the subsequent repurchase, which hinges on the seller's projection of the price of the asset. In blockchain, Shorting Attack takes on a nuanced form in Proof of Stake (PoS) based blockchains. This scenario unfolds when a malicious actor acquires a controlling stake of 51\% in the blockchain network, granting them the ability to undermine the network's integrity through a series of actions. The malicious actor initiates this strategy by selling a significant volume of the cryptocurrency short, setting off a chain reaction that compromises the blockchain's reliability. Techniques like initiating a blockchain fork and intentionally slowing down block generation are employed to introduce instability. This orchestrated chaos results in a rapid devaluation of the cryptocurrency, eroding trust. The attacker then concludes the scheme by repurchasing the cryptocurrency to stabilize the market, ultimately profiting from the decrease in value. Numerical analysis conducted by Lee et al.  \cite{lee2020short} underscores the viability of this attack strategy. Notably, due to the low stake ratio and disproportionate representation of non-malicious actors, even without wielding majority stakes, malicious actors can execute this attack to siphon significant gains, underscoring the dangerous potential of such exploits.

\textbf{Mitigation}: Ensure robust decentralization across the network, erecting a safeguard against any single miner or mining pool wielding excessive influence and potential malevolent control. A high and decentralized hash rate makes it more difficult for any single entity or group to control most of the network’s computational power. Consensus among cryptocurrency community members must be encouraged on major protocol changes and forks to avoid contentious hard forks that could undermine trust. Such checks can ensure the prevention of Shorting Attacks on blockchains. \\

\begin{figure}
\centering
\includegraphics[scale=0.32]{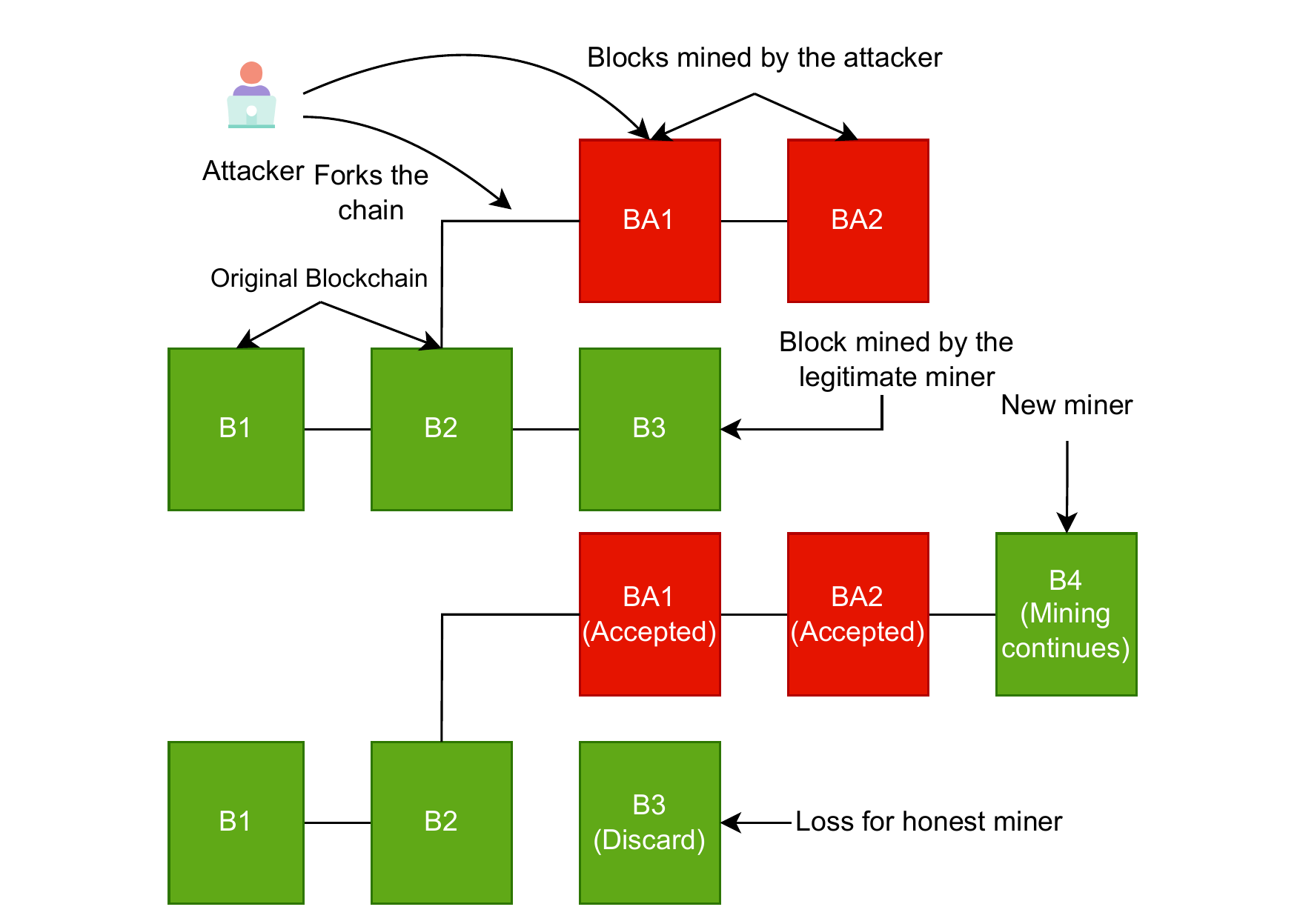}
\caption{Selfish Mining Attack. The attacker forks the blockchain and continues mining on their branch privately. Revealing their mined blocks late, the attacker's longer chain gets accepted, leading to the loss of the honest miners on the other chain.}
\label{selfishminingattack}
\end{figure}

\textbf{\textit{b)	Finality Delays \cite{neuder2021low}:}} The Ethereum 2.0 chain is susceptible to a disruptive exploit known as the Finality Delay Attack. Once a block reaches a state of finality, it becomes immune to removal from the main chain, except in specific cases when a particular network condition is met, known as "1/3-slashable." This finality typically occurs in a well-operating network within two epochs, with each epoch lasting approximately 6.4 minutes. The attacks, with a stake of 30\%, leveraging their substantial stake have the capability to defer the finality process with remarkable frequency. For instance, the attacker could delay finality by an average of three epochs per hour, corresponding to a delay of roughly 19 minutes. This orchestrated delay constitutes a potent denial-of-service (DoS) attack, affecting all the transactions dependent on finality. A recurring enactment of this strategy can potentially inflict considerable harm upon the network's vitality. This attack strategy effectively stalls the finalization of new transactions for ensuing epochs. It is essential to mention, however, that the impact is temporary. As each subsequent epoch is finalized, it leads to the retrospective finalization of all blocks from previous epochs. This process diminishes the long-term effectiveness of the disruptive scheme.

\textbf{Mitigation}: To mitigate such Finality Delay attacks, consensus mechanisms can be chosen that offer stronger finality guarantees. Some PoS-based blockchains use mechanisms like Byzantine Fault Tolerance (BFT) variants or Threshold Relay to achieve faster finality. Finality gadgets can also be used, which can help the blockchain achieve finality quicker. These are generally added as an additional layer to the blockchain. Buterin et al. \cite{buterin2017casper} introduce a new finality system, Casper, based on PoS consensus research and BFT theory. However, such additions should be made with considerable precautions since this might lead to a more complicated and risk-prone consensus system.\\

\textbf{\textit{c)	Selfish Mining Attack:}} 
The Selfish Mining Attack takes place when an attacker engineers a fork within the blockchain and starts  secretly mining an alternative branch. In this scheme, the attacker attempts to undermine the labor of legitimate miners by nullifying their concurrent branch and collecting greater rewards compared to its proportional share of mining power  \cite{selfish_mining_2023}. This attack is depicted in Figure \ref{selfishminingattack}.

Chicarino et al.  \cite{chicarino2020detection} depict the potential for brief bifurcations along a blockchain, wherein coexisting branches exist momentarily. Subsequently, nodes opt for the longest branch length and invalidate others when crafting the subsequent block. While unintentional bifurcations can occur harmlessly, leading to a delay in payments on nullified branches, this strategy relies on the presumption that no malevolent entity can gather enough mining power to introduce an alternative branch with a higher length. In alignment with the consensus algorithm, such a circumstance would prompt the adoption of the alternative branch rather than the correct one.

However, an attacker can stealthily maintain a branch exclusive to their mining efforts, through a foul block disclosure mechanism. Suppose all miners are actively mining on the xth block. If the attacker, with an advantageous head start, publishes the next block (x+1) ahead of competitors, they gain a distinct advantage. Even if the mining power of the attacker is not overwhelming, they may mine block x+1 before others, elongating the length of their branch beyond the original chain and prompting honest miners to align their efforts with the extended branch, granting the malicious actor success in the attack.

Negy et al.  \cite{negy2020selfish} introduce a more profitable intermittent Selfish Mining attack that could theoretically yield profit with a mere 37\% hash power. Their analysis suggests that the attack remains profitable even with moderate implementation difficulty. Yang et al.  \cite{yang2020ipbsm} explore a novel manifestation of this attack—intelligent bribery selfish mining (IPBSM)—where reinforcement learning aids attackers in formulating optimal strategies while interacting with the external environment.

Li et al.  \cite{li2022semi} delve into semi-selfish mining attacks with a 15\% threshold mining power, demonstrating that smaller pools face detection when attempting selfish mining tactics.

\textbf{Mitigation:} Reducing the block propagation time by quickly broadcasting newly created blocks across the network can limit the malicious actor’s attempt at selfish mining. Triggering alarms when a significant portion of the network's hash rate is being used for selfish mining can reduce the chances of this attack. Saad et al.  \cite{saad2019countering} discuss a new type of selfish mining attack with lower risk and higher reward and offer a mechanism to harness ethical mining practices to establish a novel concept for blocks within a fork known as the "truth state". They assign an expected confirmation height to each transaction to detect selfish mining activities across the network. \\

\textbf{\textit{d)	Goldfinger Attack \cite{muteba2023leveraging}:}} This is another kind of 51\% attack that particularly affects the PoW blockchain networks like blockchain. The attacker tries to destroy the asset value of the cryptocurrency they own a majority stake in. The attacker’s motivation is generally based on some incentive to outsize the cryptocurrency economy. In a market where shorting is inherently inviable, such an attack would not be a good strategy for any actor since it would degrade the value of the asset they have. Instead, a more profitable approach for miners would be to act honestly. 

However, the landscape has evolved due to the introduction of derivatives and futures markets, which have introduced the concept of a “Goldfinger” attack \cite{kroll2013economics}(named after the villain in a film whose intent is to undermine USD by destroying its gold backing). In a Bitcoin Goldfinger attack, consider a situation where an individual holds a significant role as a major Bitcoin mining entity. Due to the decreasing value of ASIC mining equipment, the diminishing effect of Moore's Law, and the emergence of new participants in the market, this individual seizes the chance to profit from shorting. They do this by amassing low-cost, out-of-the-money put options and engaging in short futures contracts. Subsequently, they carry out double-spending transactions, overwhelming network nodes, or redirecting hashing power to an alternative SHA-256 protocol. For those who wield influence within the cryptocurrency community, there could be proposals for a contentious fork, possibly with the collaboration of developers with ulterior motives. Such a fork is designed to undermine trust and involves a wide range of attacks that can threaten trustless payment systems.

\textbf{Mitigation:} Collaborating with regulatory authorities to ensure oversight and regulation of the cryptocurrency derivatives market can prevent said manipulation  \cite{goldfinger_attacks_2017}. At the same time, it should ensure that regulatory authorities do not undermine the decentralization principle of blockchain [Table 2]. Incentive structures should encourage miners to act in the network's best interest, rather than pursuing attacks. At a personal level, it is advisable to exercise caution by refraining from engaging in margin accounts across diverse platforms, alongside avoiding the utilization of stop-loss orders, even within well-liquidated exchanges. Such practices are particularly necessary due to the innate volatility of the BTCUSD pairing.

\subsection{Network Layer:}

\textit{1)	Malicious Nodes:} Blockchain networks are maintained by a decentralized collection of nodes that collaborate to validate transactions and incorporate them into the blockchain. However, malicious nodes pose a significant threat to the integrity and reliability of the blockchain. A malicious node employs unethical tactics to gain an advantage over legitimate nodes and compromise the network's operation. If a malicious actor gains control over multiple nodes or orchestrates collusion among numerous malicious nodes, they can overwhelm the blockchain network. This can lead to various attacks, disruptions, and compromises that undermine the trust and security inherent in blockchain technology.

\textbf{\textit{a)	Timejacking Attack \cite{timejacking}:}} This attack involves an attacker manipulating the timestamp information on a target node by deploying a cluster of controlled nodes to provide false time data to the victim. Timestamp accuracy is crucial for validating recently appended blocks in a blockchain network. By tampering with the victim node's timestamp, the attacker aims to disrupt this validation process. This disruption results in the rejection of blocks with timestamps exceeding a predefined time window. Consequently, the attacker can effectively isolate the victim node from the broader network, leading to a range of attacks.

There are many repercussions of a successful timejacking attack. First, it heightens the risk of double-spending transactions. Moreover, the victim node's computational resources may be siphoned off by processing fraudulent transactions. In parallel, the confirmation rates of legitimate transactions may suffer a downturn due to the manipulation. To compound matters, attackers could employ more sophisticated techniques to accelerate the time disparity between their malicious nodes and legitimate mining pools, adversely affecting the productivity of miner nodes operating on an altered chain.

\textbf{Mitigation}: To shield blockchain networks from the perils of timejacking attacks, several countermeasures can be employed \cite{timejacking}. 

•	Leveraging the network's time instead of local system time can improve timestamp accuracy. This might, however, lead to the introduction of latency in transaction validation.

•	Additionally, tightening timestamp validation parameters can effectively weed out distorted ones. 

•	Prioritizing connections with reputable nodes and monitoring network behavior during suspicious periods. 

•	Augmenting the number of confirmations required before validating transactions can mitigate the impact of fraudulent confirmations. 

•	Employing median blockchain time exclusively for validation purposes and adhering to a stringent set of trustworthy peer connections.
In conclusion, understanding the nuances of timejacking attacks and proactively implementing these countermeasures is necessary to safeguard blockchain ecosystems from this evolving threat.\\

\textbf{\textit{b)	Sybil Attack \cite{platt2021sybil}:}} Sybil attacks constitute a well-known class of threats within distributed networks. This attack involves a single malicious entity generating and overseeing numerous fabricated identities. These attacks find application in blockchain networks where they are employed to isolate a targeted victim node from the wider population of honest and authentic nodes, subsequently enabling various malicious strategies. 

In the context of blockchain, a Sybil attack unfolds with the creation of multiple deceitful identities orchestrated by a single rogue actor. The primary objective is to isolate a specific victim node from the genuine network participants. The ensuing isolation facilitates the execution of diverse forms of attacks. The attacker refrains from disseminating the transactions and mining blocks originating from the victim node to the broader network. This strategic withholding of information distorts the victim node's perception of the blockchain's state. Consequently, this skewed perspective can be exploited for ulterior motives. Additionally, the attacker can capitalize on this isolation to carry out double-spending attacks. In this scenario, only the blocks they have mined are propagated, thereby disrupting the authenticity of transactions.

\begin{figure}
\centering
\includegraphics[scale=0.3]{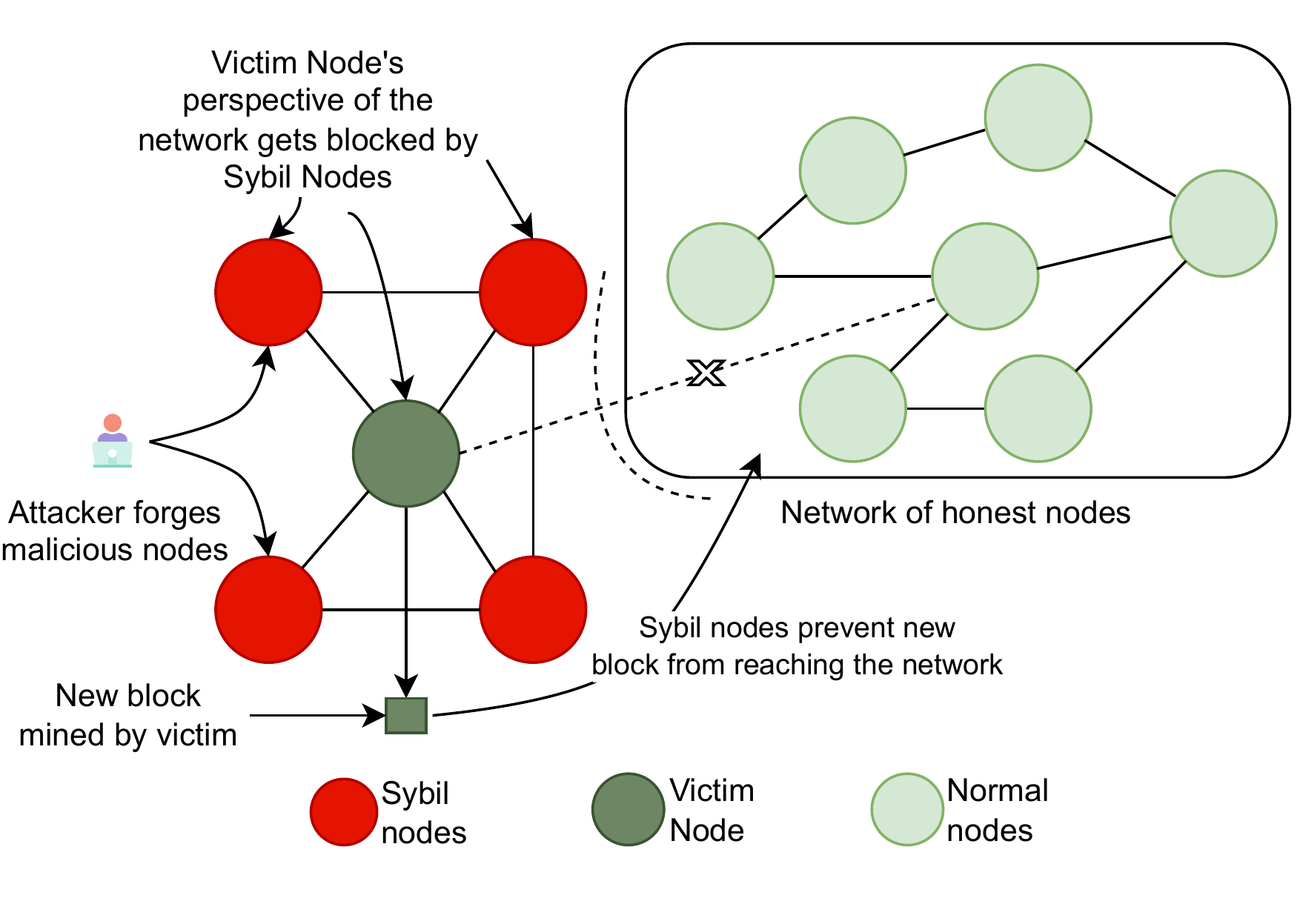}
\caption{Sybil Attack. Victim nodes get flooded with connections from the Sybil nodes of the attacker, compromising their perspective of the blockchain network.}
\label{sybilattack}
\end{figure}

Further complexity arises when the victim node is susceptible to unconfirmed payments. In this instance, the attacker can selectively filter transactions to the victim node, thereby instigating an attack on the node's operational capacity or undermining transaction confirmation rates. Such multifaceted attacks underscore the vulnerability introduced by Sybil attacks in blockchain networks. The attack is figuratively shown in Figure \ref{sybilattack}.

Sybil attacks exhibit versatility, extending to more intricate forms with the potential for evading detection. Wang et al.  \cite{wang2020truth} delve into a specific variant, the Strategic Sybil Attack, and propose an innovative defense mechanism named Truth Discovery. This approach addresses the challenges posed by Strategic Sybil Attacks within crowd-sourcing settings. The Strategic Sybil Attack involves coordinated efforts where Sybil workers strategically assign random labels to tasks. To counter this, TDSSA (Truth Discovery for Sybil Attack) introduces a Sybil score for workers, gauging the likelihood of Sybil attacker involvement. This score informs the assignments of weights to workers, guiding their task allocation. TDSSA implements probabilistic task assignments based on Sybil's scores. Workers with low scores handle high-accuracy "golden tasks," while those with higher scores are allocated tasks with more lenient accuracy requirements. This strategic assignment enhances the distinction between legitimate contributors and potential Sybil attackers, thereby boosting crowd-sourcing platform accuracy. TDSSA's efficacy surpasses existing detection methods, emphasizing its role in countering evolving Sybil attacks. 

\textbf{Mitigation}: Platt et al.  \cite{platt2021sybil} address Sybil attacks within IdAPoS consensus blockchains and provide effective countermeasures against this threat. IdAPoS employs a unique consensus mechanism where participants receive voting tokens from democratically chosen authorities, resembling a Proof of Stake (PoS) system. However, attackers can exploit this setup by allocating a significant number of tokens to attacker accounts, potentially enabling a majority attack. They propose two promising countermeasures to prevent such attacks: Temporal Normalization and TAW (Threshold-based Adaptive Weighting). These strategies aim to curb Sybil's attacks' impact by enhancing the consensus mechanism's fairness and robustness. By considering the dynamic behavior of participants and adjusting their influence accordingly, these measures strengthen the blockchain's resilience against Sybil attacks.
Douceur's work  \cite{douceur2002sybil} contributes a comprehensive insight into addressing Sybil attacks, advocating for implementing trustworthy certification. This approach relies on a centralized authority tasked with ensuring each organization possesses a single verifiable identity through issued certificates. Although a central authority is responsible for identity verification, the widespread use of this approach highlights its effectiveness in preventing Sybil attacks.

Resource testing is a robust strategy to mitigate Sybil's attacks, serving as a deterrent against the deceptive creation of multiple identities. The principle behind resource testing involves evaluating whether a group of identities possesses fewer resources than one would typically anticipate if each identity were autonomous. While this approach holds potential, it's worth noting that in various scenarios, only a handful of Sybil identities are sufficient for a successful attack.

Awerbuch et al.  \cite{awerbuch2004group} propose an innovative solution to impose recurring costs on potential Sybil attackers. They recommend incorporating Turing tests, such as Completely Automated Public Turing Tests to Tell Computers and Humans Apart (CAPTCHAs), into the authentication process. By compelling entities to solve these tests, which are easy for genuine participants but challenging for automated bots, the cost and effort of generating and managing Sybil identities increase, discouraging malicious activities.

Dragovic et al.  \cite{dragovic2003xenotrust} introduce an alternative perspective, advocating for identity certification as a deterrent mechanism against Sybil attacks. Although this certification approach lacks inherent trust, it operates as a means to impose costs on identity formation, making the creation of numerous fake identities economically unfeasible for attackers. This viewpoint underscores the broader application of economic disincentives in bolstering network security.

Gatti et al.  \cite{gatti2004sufficiently} explore the economic dimensions of network resilience and delve into the cost-effectiveness of attacks on censorship-resistant peer-to-peer networks. By leveraging an economic and game-theoretical framework, their study provides valuable insights into the financial implications of sustaining secure and censorship-resistant networks.

Entities in an application can be securely tied to a specific hardware device in a defense involving a trusted certification authority. The only way to stop a malicious actor from catching hold of a device is by manually stopping it, similar to any central authority issuing cryptographic certificates.

Swathi et al.  \cite{swathi2019preventing} present a method to keep a check on Sybil Attack. Their proposal is centered on monitoring the behavioral dynamics within a blockchain network. Each participating node assumes the role of an observer, scrutinizing the conduct and actions exhibited by nodes suspected of executing Sybil attacks. These vigilant nodes effectively defend against Sybil attacks by selectively limiting the distribution of blocks to the blockchain segment associated with a specific participant. They promptly identify and categorize suspicious nodes and expel them from the network, mitigating the potential impact of Sybil attacks and enhancing the overall robustness of the blockchain ecosystem.\\

\textbf{\textit{c)	Balance Attack \cite{natoli2016balance}:}} The Balance Attack exploits susceptibilities inherent in the GHOST (Greedy Heaviest-Observed Sub-Tree) Protocol. The GHOST Protocol, designed as a foundational framework for achieving consensus, outlines a computational process where network participants initiate from the blockchain's root block. They meticulously choose the largest subtree using a recursive algorithmic approach, capturing its root and merging it to create a coherent, shared branch. In the context of the Balance Attack, a malicious actor exploits the complexities of this protocol to create disruptions.

The Balance Attack involves a calculated disruption of communication channels between two distinct groups with comparable mining capacities. By intentionally causing a temporary disconnection, the attacker creates a situation where transactions and mining activities are simultaneously split into two separate groups — Group A and Group B. After this compartmentalization, the attacker initiates transactions within Group A while reserving mining activities for Group B. This situation persists until the accumulated structure within the block group, Group B, outbalances its counterpart within the transaction group, Group A.

The attacker tactically creates an imbalance within the GHOST Protocol's rooted structures. While certain payments are duly committed to the blockchain through Group A, the attacker ensures that they retain the power to overwrite the corresponding blocks housing these transactions. This asymmetry in structural weightage — wherein the tree within Group B surmounts that within Group A — facilitates the attacker’s potential to degrade the authenticity of committed transactions, creating an environment conducive to disruptive alterations. A point comes when the tree of transactions that the seller has visibility to becomes eclipsed by another manipulated version. The malevolent actor can exploit this structural imbalance to duplicate the payment by reallocating the same quantity of coins and triggering another payment issuance. This attack leads to a compromise of the resiliency of the original blockchain, potentially resulting in the exploitation of double spending by the malicious actor. Natoli et al. \cite{natoli2016balance} highlight the incompatibility of blockchain designs that are conducive to forking with the domain of non-public blockchains. They emphasize the significance of evaluating the architectural suitability of blockchain structures in alignment with the characteristics and objectives of specific deployment contexts.

\textbf{Mitigation}: To counter the impact of Balance Attacks, several strategies can be employed:

•	Selective Outgoing and Incoming Connections: Legitimate nodes should establish outgoing connections only with trusted nodes. By carefully choosing the nodes with which they connect, legitimate nodes can ensure that they interact with reliable participants in the network. Legitimate nodes should be cautious about accepting incoming connection requests.

•	Diverse Connection Sources: Instead of relying on a single source for connecting to nodes, legitimate nodes should diversify their connection sources, reducing the risk of a coordinated attack targeting a specific set of connections.

•	Implementing Consensus Mechanisms: Blockchain networks often use consensus mechanisms to ensure agreement among nodes about the validity of transactions. These mechanisms can detect and reject blocks proposed by malicious nodes.

•	Regular Updates and Security Measures: Keeping nodes updated with the latest security patches and measures is essential to defend against known vulnerabilities that malicious actors might exploit.

By adhering to these strategies, legitimate nodes can enhance the security of the blockchain network and minimize the impact of malicious nodes seeking to compromise the integrity of the system using Balance Attacks.\\

\textbf{\textit{d)	Eclipse Attack \cite{wust2016ethereum}:}} Eclipse attacks represent a distinct form of attack wherein the attacker tries to gain control over both incoming and outgoing connections of the victim node, effectively isolating it from the other nodes. This isolation enables the attacker to manipulate the victim's perspective of the blockchain network, drive wasteful utilization of mining power, or exploit the victim's mining capabilities for malicious ends. 

The tactical execution of an Eclipse attack involves the malicious node infiltrating the routing table of the target entity with malicious entries, initiating a sequence of actions, compelling the victim node to reinitialize and establish outbound connections with the compromised addresses residing within the table  \cite{alangot2020decentralized}  \cite{heilman2015eclipse}. Concurrently, the malicious node attempts to establish inbound connections to the victim. Gradually, a critical juncture is reached where the entirety of the victim node's connections fall under the control of the malicious entity. At this juncture, the attacker gains the ability to pass arbitrary or malicious information to the victim, thereby destabilizing its operations. Furthermore, the malicious actor can successfully replicate this method across additional honest nodes, effectively eclipsing their connections as well. In that case, they can exert dominion over the data exchange between multiple nodes. This weakness leads to the exploitation of the compromised network's integrity, potentially enabling the malicious nodes to launch attacks, including double spending attacks, which could have severe repercussions. Zhu et al.  \cite{zhu2018research} delve extensively into the intricacies of executing a typical Eclipse Attack. Wust et al.  \cite{wust2016ethereum} introduce an innovative variant of the Eclipse Attack tailored for the Ethereum Blockchain. This approach allows for the execution of the attack without monopolizing the connections of the victim node by leveraging Ethereum's architectural design centered around block transmission. 

\textbf{Mitigation}: Xu et al.  \cite{xu2020eclipsed} present the ETH-EDS model for detecting Eclipse Attacks on the Ethereum platform. This model uses a Random Forest Classification Algorithm trained on the information gathered from normal and malicious data packets, enhancing the platform's defense mechanisms. Alangot et al.  \cite{alangot2020decentralized} propose two distinct methodologies to discern potential Eclipse attack threats. One approach focuses on detecting irregular block timestamps, while the other relies on the constant communication maintained with external connections over the internet. Heilman et al.  \cite{heilman2015eclipse} meticulously quantify the resources involved in executing such attacks. Utilizing Monte Carlo simulations \cite{mooney1997monte} with real Bitcoin nodes, their experiments reveal that numerous organizations possess sufficient IP resources to launch an Eclipse Attack successfully. Furthermore, Zhang et al.  \cite{zhang2019eclipse} shed light on the utilization of Eclipse Attacks to amplify Stake-Bleeding Attacks, thereby expediting the execution of this attack variant in PoS blockchain systems. Disabling incoming connections and closely monitoring outgoing connections are recommended as countermeasures against the Eclipse Attack. \\

\textit{2)	Transaction Flooding:} Transaction flooding occurs when a targeted blockchain is overwhelmed with excessive transaction approval requests. This adversely impacts the blockchain's availability for processing legitimate transactions, affecting the network in various ways. The resulting unavailability of the blockchain can trigger denial-of-service scenarios for genuine users.\\

\textbf{\textit{a)	DDoS Attack \cite{fruhlinger2021ddos}:}} DDoS, an abbreviation for Distributed Denial of Service, represents a cyber attack wherein attackers try to render a targeted service unavailable by disrupting the operation of essential resources such as servers, applications, or specific transactions within a blockchain-based system. When the objective is to render a single system nonfunctional, the attack is termed a Denial of Service (DoS) attack. The attack is called a Distributed Denial of Service (DDoS) attack if the intention is to disrupt multiple systems concurrently. These attacks operate by flooding the target system with an overwhelming number of fraudulent requests. For instance, consider a web page that is flooded with a barrage of requests to serve a web application; the resultant surge in requests can potentially lead to the crash of the web app due to the overload. Likewise, a database encountering an abrupt surge in queries might become unresponsive due to the excessive load. The repercussions of such attacks range from inconveniencing users by hampering their service experience to the complete shutdown of entire businesses.

DDoS attacks encompass two primary categories \cite{fruhlinger2021ddos}. The initial category is volume-based attacks, which deploy a multitude of fake transactions to disrupt the targeted blockchain system. The extent of such attacks is quantified by measuring bits per second (BPS). The second category contains protocol or network-layer attacks, characterized by transmitting an extensive volume of packets to the chosen victim within the blockchain infrastructure. The magnitude of this type of attack is gauged using the measurement of packets per second (PPS). 

A comprehensive game-theoretical analysis of these DDoS attacks within the context of Bitcoin is conducted by Johnson et al.  \cite{johnson2014game}. This study delves into the intricate trade-offs that miners encounter when contemplating the increase in their computational power while simultaneously conducting a Distributed Denial of Service (DDoS) attack through a mining pool. Notably, the vulnerability to such attacks is higher for larger mining pools. The incentive for larger mining pools to carry out these attacks increases as the rewards outweigh those for smaller pools.

Luo et al.  \cite{luo2021preventing} undertake a comprehensive analysis of DDoS attacks on the Bitcoin memory pool and outline the potential repercussions these attacks pose on various aspects of the blockchain. They introduce a defense mechanism termed the Dynamic Fee Threshold Mechanism. In the domain of smart contracts, Kumar et al.  \cite{kumar2021distributed} present a decentralized architecture rooted in fog computing to identify DDoS attacks. Their approach aims to enhance detection through a fog computing framework.

\textbf{Mitigation}: Chaganti et al.  \cite{chaganti2022csomprehensive} describe various state-of-the-art DDoS attacks and the existing mitigation solutions against them. The authors also categorize these attacks based on the vulnerability of Blockchain-based distributed systems these attacks target. They also discuss how most solutions focus on keeping a record of the malicious addresses in the transactions and distributing these IP addresses across the network. Jia et al.  \cite{jia2020anti} introduce a distributed anti-D chain detection system characterized by hybrid ensemble learning and a virtual reality parallel anti-DDoS chain design philosophy. Using machine learning classifiers such as Random Forest and AdaBoost, supplemented by lightweight algorithmic classifiers like ID3, their methodology demonstrates improved accuracy in identifying the aspects associated with DDoS attacks. Substantial experimentation validates the enhanced performance of this detection methodology across multiple essential indicators.\\

\textbf{\textit{b)	Spam Attack \cite{spamming}:}} In layer 1 public blockchains, including Solana, Near, Avalanche, and Harmony, which compete with Ethereum, the appeal lies in their promise of offering higher performance, minimal latency, and cost-effectiveness. However, they are susceptible to a persistent security threat known as transaction spamming, or Tx spamming. Spam attacks essentially function as Denial-of-Service (DoS) attacks, capable of causing an abnormal increase in on-chain traffic and an accelerated growth in the blockchain database size. This change can happen either gradually or over an extended timeframe, resulting in prolonged delays in computing regular transactions. Sometimes, validator nodes may lose consensus, rendering them incapable of producing valid blocks, leading to chain halts.

\textbf{Mitigation}: One effective measure to counteract such attacks on blockchain networks is to increase the minimum transaction fee \cite{spamming} marginally. This adjustment, while not significantly impacting regular transactions, can deter spam attackers, particularly given the high volume they typically operate with. An illustrative example of this approach occurred in September 2021 when Polygon faced a spam onslaught orchestrated by two arbitrage trading bots, generating a staggering two million transactions daily due to the low 1 gwei gas price. In response, the Polygon team raised the minimum transaction fee from 1 gwei to 30 gwei, resulting in a notable 75\% reduction in the volume of spam transactions  \cite{the_money_making_machine}.\\

\textbf{\textit{c)	Dust Attack \cite{dusting}:}} In cryptocurrencies, dust refers to minuscule amounts of digital currency distributed across a multitude of wallet addresses. Typically, dust encompasses quantities of cryptocurrency equal to or less than the associated transaction fees. Additionally, dust may arise as a residue of trading activities, attributable to rounding errors or transaction fees, gradually accumulating over time. While this nominal sum may not be directly tradable, it can be converted into the native token of the exchange platform.

\begin{figure}
\centering
\includegraphics[scale=0.4]{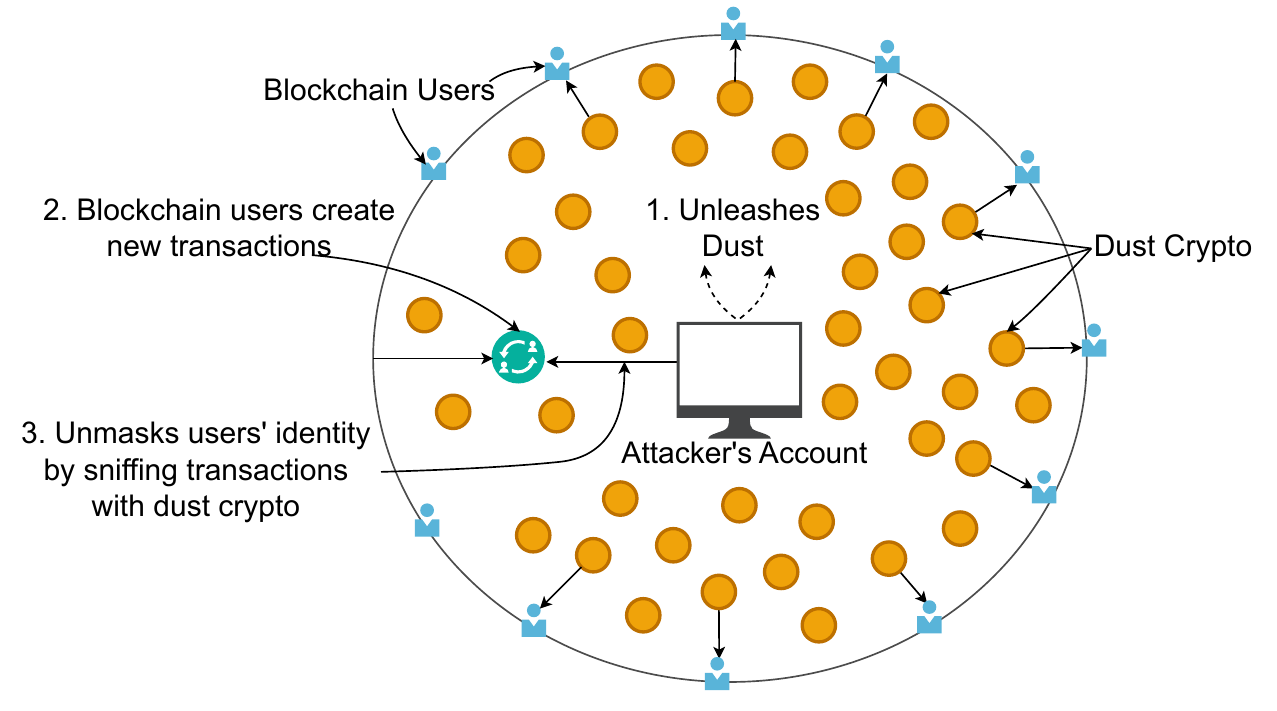}
\caption{Dust Attack. The steps are numbered 1 to 3. The attacker releases a lot of dust on cryptocurrencies in the network. If the victims accept the dust, the attacker can track their transactions containing the dust and unmast their identities.}
\label{dustattack}
\end{figure}

A Dust Attack occurs when attackers disperse such minuscule amounts of dust across various wallet addresses scattered throughout blockchain networks. Within blockchain networks, all transactions are transparent and visible to all participants, permitting tracking of user activities by following the transaction history of specific addresses. When attackers dispatch dust to cryptocurrency wallets, their objective is to breach the privacy of wallet owners, shadowing the movement of funds from one address to another. The attackers try to link the target's address with other addresses, potentially unmasking the victim's identity through off-blockchain hacking. Subsequently, they may launch many attacks, including phishing schemes, cyber extortion threats, blackmail, or identity theft, to profit from the victim's vulnerability. Figure \ref{dustattack} illustrates the Dust Attack.

These attackers exploit that cryptocurrency users often overlook receiving meager digital currency amounts in their wallet addresses. Notably, transaction patterns are traceable, which could lead to the identification of wallet owners. For a dust attack to achieve its intended impact, the wallet owner must merge the crypto dust with other funds within the same wallet and employ it in subsequent transactions. By incorporating a small amount of cryptocurrency into these transactions, the target of the attack may inadvertently transfer the dust to an off-blockchain centralized entity. As these centralized platforms must adhere to Know Your Customer (KYC) regulations, they retain the victim's data, rendering them susceptible to off-blockchain attacks. Analogous to how we employ small changes in conventional financial transactions, crypto dust from multiple addresses can be utilized in other transactions. Detecting the origin of funds from the dust attack transaction, attackers can harness advanced technological tools to trace a trail, ultimately revealing the victim's identity. 

\textbf{Mitigation}: Due to the escalating transaction fees in most blockchain networks, executing dust attacks has become progressively challenging. Wallet owners are encouraged to employ privacy-enhancing tools such as TOR or VPNs to secure their anonymity and security \cite{dusting}. Utilizing a Hierarchical Deterministic (HD) wallet, which automatically generates a fresh address for each transaction, can further prevent hackers from attempting to trace transaction histories. Wang et al. \cite{wang2018anti} propose a novel method for identifying Dust Attacks by recognizing such attacks using Gaussian distribution.  

\subsection{Data Layer:}

\textit{1)	Weak Credentials:} The susceptibility of weak credentials arises when users within a blockchain network employ inadequate authentication, leaving their accounts vulnerable to exploitation by attackers who can impersonate the user and execute damaging actions.\\

\begin{figure}
\centering
\includegraphics[width=1\linewidth]{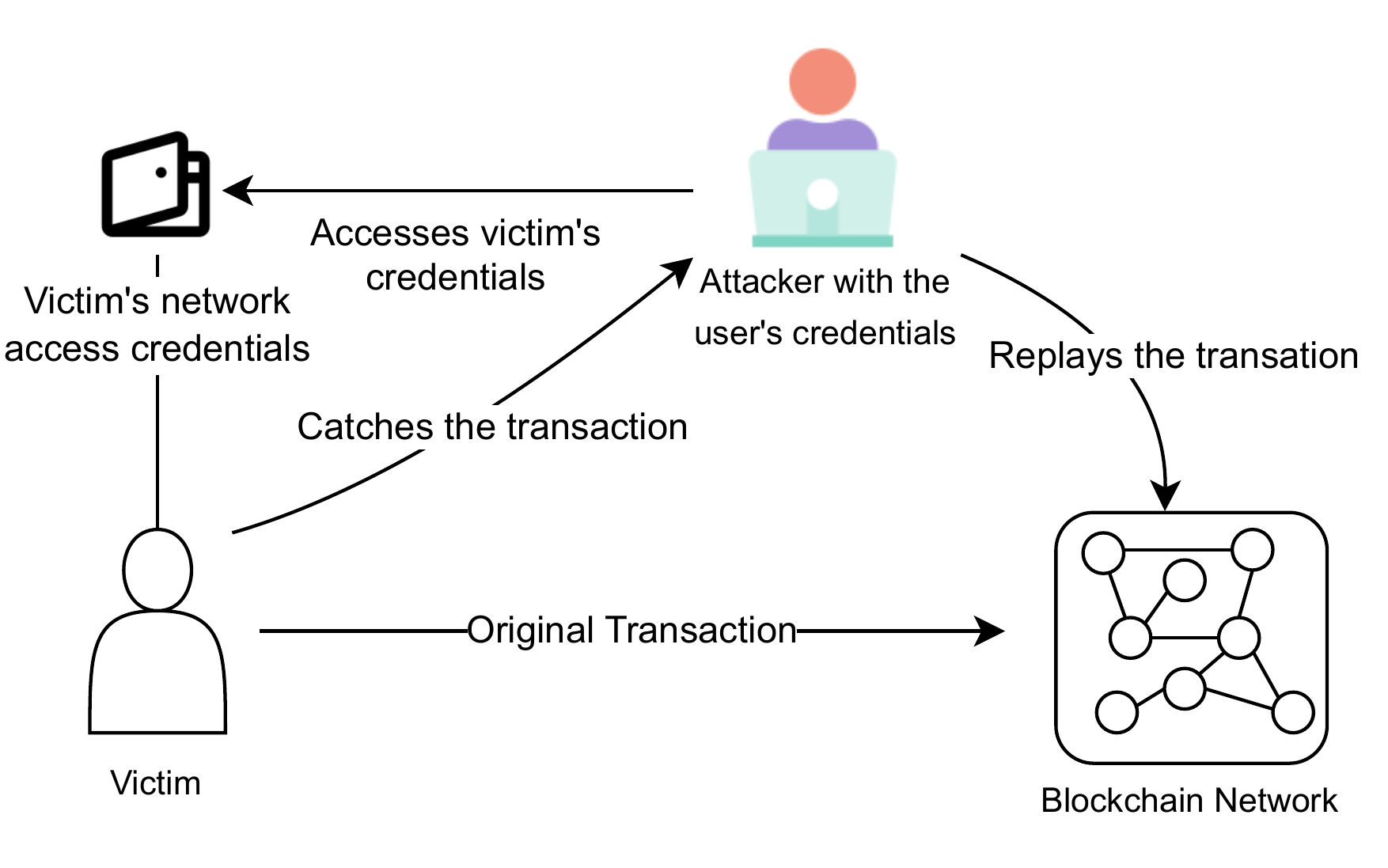}
\caption{Replay Attack. The attacker sniffs the transaction a victim is submitting to the blockchain. They then replay the transaction using the victim's credentials.}
\label{relayattack}
\end{figure}

\textbf{\textit{a)	Replay Attack \cite{anita2019blockchain}:}} A replay attack constitutes a form of attack wherein an actor falsifies communication between two authentic entities to gain unauthorized access. In this attack, the attacker intercepts a genuine transaction and subsequently reproduces it, transmitting the duplicated transaction across the distributed network. The feasibility of this attack arises from the malicious actor possessing authentic credentials, allowing them to access information within the network. Due to the credentials' authenticity, security algorithms do not flag this attack as malicious, treating it as a routine data exchange. Figure \ref{relayattack} displays the Replay Attack.

This attack aligns seamlessly with the architecture's structure, relying primarily on valid credentials and precise timing for successful execution. The attacker can adeptly command network access credentials, effectively assuming the identity of a legitimate user and thereby gaining access to the user's complete transaction history. This condition serves as both a necessary and sufficient requirement for the execution of a replay attack, bearing potential catastrophic consequences.

Within the blockchain context, replay attacks can escalate to massive proportions. This vulnerability is amplified due to the diminishing computing power of legacy blockchains, creating an opportunity for a 51\% attack. This scenario enables the injection of new transactions into the blockchain, potentially rendering it inoperable if the computational power of the attacker surpasses that of the legitimate participants.

\textbf{Mitigation}: Countermeasures against replay attacks within the blockchain domain can be categorized into two primary classes \cite{replay}. The first, strong replay protection, involves introducing unique markers after a hard fork generates a new ledger, ensuring that transactions conducted on the novel ledger are incompatible with the original one and vice versa. The second approach, opt-in replay protection, mandates participants to modify their transactions to prevent replication. This approach is particularly relevant when hard forks aim to enhance the existing blockchain rather than create a network split. Employing key-pair-based exchange mechanisms can shield participants against these attacks.

Timestamping all transactions within mined blocks offers another defense against replay attacks. By associating transaction packets with the time of their occurrence, the attacker's ability to resend transactions after a certain time frame is restricted. Similarly, employing one-time passwords for transactions serves as an additional layer of protection against replay attacks.\\

\textbf{\textit{b)	Dictionary Attack \cite{pinkas2002securing}:}} Dictionary Attacks represent attacks in which the attacker employs hash values of frequently used passwords, such as "password123," to compromise the target's cryptographic hash. This attack involves attempting various hash values derived from commonly employed passwords. By converting plain text passwords into cryptographic hashes, attackers seek to identify wallet credentials and gain unauthorized access to the target's resources.

\textbf{Mitigation}: Pinkas et al.  \cite{pinkas2002securing} examine the dictionary attack and propose several countermeasures that users can implement to protect their passwords against malicious actors. \\

\textbf{\textit{c)	Cryptojacking Attack:}} Cryptojacking  \cite{eskandari2018first} is an emerging threat that infiltrates computing devices, using their computational resources to mine cryptocurrencies secretly. Cryptocurrencies, such as Bitcoin and altcoins, are digital or virtual assets relying on the decentralized ledger of blockchain. Miners in a PoW solve complex mathematical puzzles, facilitate transactions, and contribute to the blockchain's integrity. Notably, the energy-intensive nature of PoW mining sets it apart, with the Bitcoin network alone consuming over 73TWh of electricity annually.

Cryptojacking emerges from the desire for cryptocurrency mining rewards without incurring substantial operational costs. This practice allows attackers, or cryptojackers, to mine cryptocurrency while evading the financial overhead associated with legitimate mining operations. Monero \cite{akcora2022blockchain}, a privacy-focused cryptocurrency, remains a preferred choice for cybercriminals due to its untraceable nature. Cybercriminals employ two primary methods to engage victim’s devices in cryptocurrency mining:

•	Malicious Links: Victims are deceived into clicking malicious links within emails, leading to the execution of crypto mining scripts on their computers.

•	Browser-Based Attacks: Hackers inject JavaScript code into websites or online ads. This code auto-executes when loaded in victims' browsers, initiating background mining operations. Both methods aim to leverage computational resources covertly.

While cryptojacking scripts do not inflict direct damage on computers or data, they deplete processing resources. For individual users, the consequence may be slow device performance. However, organizations with numerous cryptojacked systems incur tangible costs, including increased IT workload, heightened electricity expenses, and potential network infections. Some cryptojacking scripts possess worm-like capabilities, propagating across networks and infecting additional devices and servers, complicating detection and eradication efforts. Moreover, these scripts may actively disable competing crypto-mining malware on infected devices. Mobile devices are susceptible to cryptojacking through tactics paralleling those employed against desktop systems. Trojans concealed within downloaded apps or redirects to infected sites on users' phones enable attackers to infiltrate mobile devices. Although individual mobile devices have limited processing power, collective compromises can yield substantial rewards for cryptojackers.

\textbf{Mitigation}: To prevent cryptojacking, users are recommended to use reputable anti-virus and anti-malware software  \cite{cryptojacking} on their devices. Browser extensions or add-ons blocking online ads and cryptojacking scripts are also popular. Strong and unique passwords are recommended to prevent unauthorized access to their devices and accounts.\\

\textbf{\textit{d)	Packet Sniffing Attack \cite{zhu2018security}:}} A packet sniffing attack refers to the unauthorized interception and inspection of data packets exchanged between nodes in a blockchain network. These packets contain information related to transactions, smart contracts, and other blockchain-related data. The captured data packets may include sensitive information such as transaction details, wallet addresses, private keys, and other data associated with blockchain transactions. The attacker aims to gain access to this information for other attacks. If an attacker successfully intercepts and analyzes blockchain data packets, they can potentially compromise the privacy and security of blockchain users. 

\textbf{Mitigation}: The risk of packet sniffing attacks may vary depending on whether the blockchain is public or private. Public blockchains, such as Bitcoin and Ethereum, have openly accessible network traffic, making them potentially more vulnerable to packet sniffing by anyone with network access. Private or permissioned blockchains typically have stricter communication protocols, implement firewall rules, ensure the physical security of network infrastructure, and monitor network traffic for suspicious activities.\\

\textit{2)	Key Generation Vulnerabilities: } To establish data authenticity within blockchains, various cryptographic tools and algorithms are utilized to generate digital signatures. These tools pertain to generating and managing cryptographic keys, which are essential for secure transactions, identity management, and data encryption in blockchain applications. Over the past several years, many improvements have been made to make these cryptographic algorithms more robust. However, several vulnerabilities, like entropy accumulation, which leads to insufficient randomness during key generation, and faulty key storage and exposure, plague these algorithms. Some examples of how these vulnerabilities are plaguing blockchains are mentioned below.

\textbf{\textit{a)	Lower Entropy in ECDSA \cite{hughes2021badrandom}:}}  The Elliptic Curve Digital Signature Algorithm (ECDSA), employed by Bitcoin to generate private-public key pairs, has been criticized for having inadequate entropy. This deficiency in entropy can lead to imprecise information being present in multiple signatures, potentially compromising their reliability and effectiveness.\\

\textbf{\textit{b)	Private Key Theft in Ed25519:}}Ed25519, a prominent digital signature algorithm, has gained widespread recognition within the cryptographic community, particularly in the context of cryptocurrency and blockchain platforms. Its preference over ECDSA can be attributed to its openness, enhanced security, and superior speed. In blockchain technology, Ed25519 has gained massive traction.

\begin{table*}[]
\scriptsize
\begin{tabular}{|c|c|c|cc|c|c|}
\hline
\textbf{Layer} &\textbf{Vulnerability} & \textbf{Attack}  & \multicolumn{2}{c|}{\textbf{Mitigations}}  & \textbf{Advantages} & \textbf{Disadvantages}  \\ \hline
\multirow{3}{*}{\begin{tabular}[c]{@{}c@{}}Application\\ Layer\end{tabular}} & \multirow{3}{*}{\begin{tabular}[c]{@{}c@{}}0-Confirmed  \\ Transactions\end{tabular}}  & Race Attack  & \multicolumn{2}{c|}{\multirow{3}{*}{\begin{tabular}[c]{@{}c@{}}Await   transaction confirmation from multiple \\ trusted miners before validating a transaction\end{tabular}}}  & \multirow{3}{*}{\begin{tabular}[c]{@{}c@{}}Ensures that goods and services \\are being extended\\ against verified transactions\end{tabular}} & \multirow{3}{*}{\begin{tabular}[c]{@{}c@{}}Higher turn around time\\ for transaction acceptance\end{tabular}} \\ \cline{3-3}
& & Finney Attack & \multicolumn{2}{c|}{} & &  \\ \cline{3-3}
& & Vector76 Attack & \multicolumn{2}{c|}{} & & \\ \hline
\multirow{7}{*}{\begin{tabular}[c]{@{}c@{}}\\\\\\\\\\\\\\\\\\\\\\\\\\Contract\\Layer\end{tabular}} & \begin{tabular}[c]{@{}c@{}}Faulty Access \\ Specifier\end{tabular} & Default Visibilities & \multicolumn{2}{c|}{\begin{tabular}[c]{@{}c@{}}Modifying the default access specifier from public \\to private in Solidity\end{tabular}} & \begin{tabular}[c]{@{}c@{}}Functions missing the access \\specifier will not be       vulnerable \\to malicious  external calls\end{tabular} & \begin{tabular}[c]{@{}c@{}}Private contracts have higher\\ gas consumption for inter\\ contract  communication cases\end{tabular} \\ \cline{2-7} 
 & \multirow{2}{*}{\begin{tabular}[c]{@{}c@{}}\\Unauthorized \\ Input\end{tabular}} & \begin{tabular}[c]{@{}c@{}}Overflow and\\Underflow\end{tabular} & \multicolumn{2}{c|}{{\begin{tabular}[c]{@{}c@{}}1. Properly check the inputs to the smart contract\end{tabular}}}  & {\begin{tabular}[c]{@{}c@{}}1. Smart contract will become\\  more robust to malicious\\  inputs from external \\users\end{tabular}}  & \multirow{2}{*}{-} \\ \cline{3-3}
 & & \begin{tabular}[c]{@{}c@{}}Short Address \\ Attack\end{tabular} & \multicolumn{2}{c|}{\begin{tabular}[c]{@{}c@{}} 2. Using standard libraries for code audits \end{tabular}}  &  \begin{tabular}[c]{@{}c@{}}2. Standard libraries will tackle\\ such values properly\end{tabular}  &  \\ \cline{2-7} 
 & \multirow{4}{*}{\begin{tabular}[c]{@{}c@{}}\\\\\\\\\\\\\\Smart\\ Contract \\ Bugs\end{tabular}} & Reentrancy Attack  & \multicolumn{2}{c|}{\begin{tabular}[c]{@{}c@{}}1. Updating  balance before transaction initiation\\\\2. Using mutex locks in smart contracts calling \\external functions\end{tabular}}  & \begin{tabular}[c]{@{}c@{}}1. Will keep the balance in the\\smart contract in check,\\preventing circular calls\\ \\2. Will stop multiple executions \\of the same function\end{tabular}  & \begin{tabular}[c]{@{}c@{}}1. Such updation can\\lead to race conditions\\\\2. Can lead to higher \\ gas costs and deadlocks \end{tabular} \\ \cline{3-7} 
 &  & \begin{tabular}[c]{@{}c@{}}Tx. Origin \\ Vulnerability\end{tabular} & \multicolumn{2}{c|}{\begin{tabular}[c]{@{}c@{}}Use "msg.sender" instead of "tx.origin" for caller\\authentication\end{tabular}}   & \begin{tabular}[c]{@{}c@{}}The attacker will not be \\able to pose as the owner\\of the vulnerable smart\\contract\end{tabular}  & \begin{tabular}[c]{@{}c@{}}Relying on msg.sender  can \\lead to Reentrancy Attack \\ if not handled properly\end{tabular}\\ \cline{3-7} 
 & & \begin{tabular}[c]{@{}c@{}}Self Destruct \\ Attack\end{tabular} & \multicolumn{2}{c|}{\begin{tabular}[c]{@{}c@{}}1.   Self destruct condition should have\\ no dependence on  balance in the contract\\\\2. Ensure that ether transfer during self destruction\\is sent to trusted address\end{tabular}}  & \begin{tabular}[c]{@{}c@{}}1. Attacker can not invoke\\the self-destruct\\ by manipulating the\\balance\\\\ 2. In case of self-destruct,\\ether will not\\be transferred to the\\ attacker's account\end{tabular}  & - \\ \cline{3-7} 
 & & Gasless Send & \multicolumn{2}{c|}{\begin{tabular}[c]{@{}c@{}}1. Instead of using "send" function, use "transfer" \\or "call.value"\\\\2. Perform thorough gas estimation before\\executing transactions\end{tabular}} & \begin{tabular}[c]{@{}c@{}}1.   Specifying the gas amount\\ using these  functions\\ will limit the attack\\\\2. Attacker will not be able\\to force-execute transactions\\with arbitrary gas value\end{tabular}  & \begin{tabular}[c]{@{}c@{}}1.Transfer and call.value \\ functions have lesser  control\\ over gas,  leading to \\ operational inefficiency\\\\ 2. Will lead to wastage \\ of significant resources \\ in case of failed transactions\end{tabular} \\ \hline
 \multirow{8}{*}{\begin{tabular}[c]{@{}c@{}}\\\\\\\\\\\\\\\\\\\\\\\\\\\\Consensus\\ Layer\end{tabular}} & \multirow{4}{*} 
 {\begin{tabular}[c]{@{}c@{}}\\\\\\\\\\\\Blockchain   \\ Centralization\end{tabular}} & Shorting Attack & \multicolumn{1}{c|}{-} & 
 \multirow{4}{*}{\begin{tabular}[c]{@{}c@{}}\\\\\\\\4. Improve\\blockchain\\decentralization\\using a\\high and\\decentralized\\hash 
 rate.\end{tabular}} & {\begin{tabular}[c]{@{}c@{}}\\1. Malicious activities in the\\ cryptocurrency derivatives\\ market will be put to\\ check \end{tabular}} & \multirow{4}{*}{\begin{tabular}[c]{@{}c@{}}1. Introduction of regulatory \\ authorities will damage 
 \\ the decentralization property \\ of blockchain\\  \\ 3. Consensus mechanisms with\\ lower block propagation time \\ generally have higher risks of \\ forks and latency requirements\\ \\ 4. Higher hash rates will lead\\ to more resource intensive\\mining\end{tabular}} \\ \cline{3-4}
 & & \begin{tabular}[c]{@{}c@{}}Goldfinger \\ Attack\end{tabular} & \multicolumn{1}{c|}{\begin{tabular}[c]{@{}c@{}}1. Regulatory authorities should \\ ensure oversight \\ of derivatives market\\ \\2. Refrain from engaging \\ in margin accounts and \\ avoid using stop loss orders\end{tabular}} & & {\begin{tabular}[c]{@{}c@{}}\\2. Goldfinger attack will not\\ take place if  margin\\ accounts are avoided \\\\ 3. Lower block \\propagation time\\ will limit the attacker's \\ chances to manipulate\\ the blockchain \end{tabular}} & \\ \cline{3-4}
 & & Selfish Mining & \multicolumn{1}{c|}{\multirow{2}{*}{\begin{tabular}[c]{@{}c@{}}\\3. Choose consensus \\mechanisms that reduce \\the block propogation time\end{tabular}}} & & & \\ \cline{3-3}
 & & Finality Delays & \multicolumn{1}{c|}{} & & {\begin{tabular}[c]{@{}c@{}} 4. Decentralization of the\\ blockchain will make it\\ difficult  for attacker \\to execute these attacks\\\\  \end{tabular}} & \\ \cline{2-7} 
 & \multirow{4}{*}{\begin{tabular}[c]{@{}c@{}}\\\\\\\\Blockchain \\ Forkability\end{tabular}} & \begin{tabular}[c]{@{}c@{}}Malicious \\ Reorgs\end{tabular} & \multicolumn{1}{c|}{-} & \multirow{4}{*}{\begin{tabular}[c]{@{}c@{}}\\2. Random \\ allocation\\ of miners to\\ different\\ branches\\\\ 3. Set threshold\\ limits\\ for mining\\  pools\\ \end{tabular}} & \multirow{4}{*}{\begin{tabular}[c]{@{}c@{}}1.   Loss of security deposit\\ will dissuade  attackers to\\ act maliciously\\\\ 2. Attackers will not be \\able to mine on their preferred \\chain to execute the \\attack\\\\3. Will limit attacker's ability\\ to take  advantage of forks\end{tabular}} & \multirow{4}{*}{\begin{tabular}[c]{@{}c@{}}1. Security deposits for \\ block validation will deter \\ more  people to participate \\ in block   validation\\ \\2. Will lead to economic \\ disincentives and inefficient \\ resource utilization for \\ miners on lower reward\\  branches\\\end{tabular}} \\ \cline{3-4}
 & & FAW Attack & \multicolumn{1}{c|}{-} & & & \\ \cline{3-4}
 & & Stalker Attack & \multicolumn{1}{c|}{-} & & & \\ \cline{3-4}
 & & \begin{tabular}[c]{@{}c@{}}Nothing-at-Stake \\ Attack\end{tabular} & \multicolumn{1}{c|}{\begin{tabular}[c]{@{}c@{}}\\\\\\1. Require the validators \\ to submit a security-deposit \\ for block validation\\\\\\\end{tabular}} & & & \\ \hline
\end{tabular}
\caption*{Table 2: Advantages and Disadvantages of various mitigations(continued)}
\end{table*}

\begin{table*}[]
\scriptsize
\begin{tabular}{|c|c|c|cc|c|c|}
\hline
\textbf{Layer} & \textbf{Vulnerability} & \textbf{Attack} & \multicolumn{2}{c|}{\textbf{Mitigations}} & \textbf{Advantages} & \textbf{Disadvantages} \\ \hline
\multirow{6}{*}{\begin{tabular}[c]{@{}c@{}}\\\\\\\\\\\\\\\\\\\\\\\\Network\\ Layer\end{tabular}} & \multirow{3}{*}{\begin{tabular}[c]{@{}c@{}}Transaction\\ Flooding\end{tabular}} & DDoS Attack & \multicolumn{1}{c|}{-} & \multirow{3}{*}{\begin{tabular}[c]{@{}c@{}}2. Increase\\ minimum\\ transaction fee\end{tabular}} &{\begin{tabular}[c]{@{}c@{}}1. Fresh addresses generated \\ by HD wallets will thwart hackers in \\ their attempts to trace transaction \\histories\\ \end{tabular}} & {\begin{tabular}[c]{@{}c@{}}1. HD wallets are more\\ complex and have\\ privacy concerns due to \\reliance on single\\  master key for all \\address generation\\ \end{tabular}} \\ \cline{3-4}
& & Spam Attack & \multicolumn{1}{c|}{-} & & & \\ \cline{3-4}
& & Dust Attack & \multicolumn{1}{c|}{ \begin{tabular}[c]{@{}c@{}}1. Utilizing a Hierarchical \\ Deterministic (HD)wallet\end{tabular}} & & {\begin{tabular}[c]{@{}c@{}}2. Will significantly increase the\\ cost of such attacks\end{tabular}} & {\begin{tabular}[c]{@{}c@{}}2. Will increase transaction \\fee for normal transactions\end{tabular}} \\ \cline{2-7} 
& \multirow{3}{*}{\begin{tabular}[c]{@{}c@{}}\\\\\\\\\\\\Unverified \\ Connections \\ to Nodes\end{tabular}} & \begin{tabular}[c]{@{}c@{}}Timejacking   \\ Attack\end{tabular} & \multicolumn{1}{c|}{\begin{tabular}[c]{@{}c@{}}1. Use network's time instead\\ of local system time\\\\ 2. Tightening validation  parameters\\ for timestamps\end{tabular}} & \multirow{3}{*}{\begin{tabular}[c]{@{}c@{}}\\\\\\\\\\6.   Properly \\ auditing \\ incoming and \\ outgoing \\ connections \\ in the  network\end{tabular}} & {\begin{tabular}[c]{@{}c@{}}1. Attacker will not be able to \\manipulate timestamp if network's \\ time is used\\ \end{tabular}} & \multirow{3}{*}{\begin{tabular}[c]{@{}c@{}}\\\\\\1.   Network time\\ synchronization may\\ introduce delays in\\ timestamp accuracy\\ \\ 3, 4. Will damage the\\ decentralization principle \\of blockchains\\ \\ 5. Will add operational\\ overhead for honest nodes\end{tabular}} \\ \cline{3-4}
& & Balance Attack & \multicolumn{1}{c|}{-} & & {\begin{tabular}[c]{@{}c@{}} 2. Any attempt by attacker will be\\ dealt with using proper validation \\ of timestamps\\ \end{tabular}} & \\ \cline{3-4}
& & Sybil Attack & \multicolumn{1}{c|}{\begin{tabular}[c]{@{}c@{}}\\3.   Centralized entity ensuring \\ each organization having \\ identifiable identity\\\\ 4. Evaluating whether a \\ group of identities possesses fewer \\ resources than a typical autonomous\\ one.\\\\5. Incorporating Turing tests in\\ authentication process\\ \\ \end{tabular}} & & {\begin{tabular}[c]{@{}c@{}} 3, 4, 5. Illicit nodes will be flagged\\ and brought down with these \\ validation checks\\   \\   6. It can be ensured that the\\ connecting nodes will not try to \\manipulate the blockchain network \\for your node\end{tabular}} & \\ \hline
\multirow{4}{*}{\begin{tabular}[c]{@{}c@{}}\\\\\\\\Data\\ Layer\end{tabular}} & \multirow{4}{*}{\begin{tabular}[c]{@{}c@{}}\\\\\\\\Poor\\ Encryption\end{tabular}} & Replay Attack & \multicolumn{2}{c|}{\begin{tabular}[c]{@{}c@{}}1. Adding unique markers on transactions after \\ a hard fork generates a new ledger\\\\ 2. Timestamping all transactions among \\mined blocks\end{tabular}} & \begin{tabular}[c]{@{}c@{}}1, 2. Replay attacks will be spotted\\ easily with the unique \\markers and proper timestamps\end{tabular} & \begin{tabular}[c]{@{}c@{}}1, 2. Can lead to additional \\ gas cost for\\ transactions\end{tabular} \\ \cline{3-7} 
& & \begin{tabular}[c]{@{}c@{}}Dictionary \\ Attack\end{tabular} & \multicolumn{2}{c|}{\begin{tabular}[c]{@{}c@{}}Using random and secure passwords \\ for securing wallets\end{tabular}} & \begin{tabular}[c]{@{}c@{}}Will make it difficult to\\ brute force the password\end{tabular} & - \\ \cline{3-7} 
& & Cryptojacking & \multicolumn{2}{c|}{\begin{tabular}[c]{@{}c@{}}Using third party services like anti-virus \\ softwares, browser add-ons\end{tabular}} & \begin{tabular}[c]{@{}c@{}}Will stop malicious users\\ to highjack computing power\end{tabular} & - \\ \cline{3-7} 
& & Packet Sniffing & \multicolumn{2}{c|}{\begin{tabular}[c]{@{}c@{}}Having   stricted communication protocols, \\ firewall rules for data packets in the   network\end{tabular}} & \begin{tabular}[c]{@{}c@{}}Data packets will be hard to\\ track and read\end{tabular} & \begin{tabular}[c]{@{}c@{}}These might lead to \\ higher resource requirements\\  and introduction of latency\\ in high-traffic environments\end{tabular} \\ \hline
\end{tabular}
\caption{Advantages and Disadvantages of various mitigations(continued from previous page)}
\label{summarytable}
\end{table*}

One of the primary advantages of Ed25519 over ECDSA is its deterministic signature generation process. Unlike ECDSA, EdDSA signatures do not necessitate access to a secure Random Number Generator (RNG) during transaction signing. This becomes especially advantageous where a significant proportion of devices, such as laptops and IoT devices, lack a robust source of entropy or employ a weak RNG mechanism. The standard Ed25519 message signing procedure provides the algorithm with a message and a private key. The function employs the private key to compute the corresponding public key and subsequently signs the message. Some libraries offer a modified version of the message signing function, which accepts the pre-computed public key as an input parameter. While this implementation offers certain benefits, including improved efficiency by avoiding repetitive public key derivation, it inadvertently introduces a security vulnerability within the library. Notably, Chalkias  \cite{cryptography} identified instances where certain libraries permitted arbitrary public keys as inputs without verifying their correspondence to the provided private key. This oversight opens a potential avenue for attackers to exploit the signing function as an Oracle, enabling cryptanalysis and the potential revelation of sensitive information. For instance, in specific applications where key generation fails or a cleanup process deletes the private key for a user, there exists a brief window of opportunity during which the database retains the old <userID, pubKeyOld> pair. This scenario creates a vulnerability susceptible to race condition attacks before updating the database with the new public key. This vulnerability shows the importance of thorough security assessments.

\textbf{Mitigation}: In response to these identified vulnerabilities, several libraries have taken proactive measures by introducing fixes for existing vulnerabilities or implementing additional checks to ensure the correspondence between stored public and private keys. These security enhancements aim to fortify the integrity of Ed25519 implementations within the blockchain and cryptographic ecosystems. Other encryption schemes can also be considered as a replacement for EdDSA. In  \cite{bisogni2021ecb2}, the authors propose a novel encryption scheme that uses the user's face as a biometric key, encoded using FaceNet.

We summarize the findings of all discussed vulnerabilities, attacks, and mitigations on each layer in Table \ref{summarytable}. Figure \ref{countermeasures} serves as a reference, presenting the prevailing solutions documented within the literature that serve to counter these attacks. We also mention the possible mitigation steps along with the attacks. However, all such mitigations have advantages and disadvantages, highlighted in Table \ref{summarytable}. Before implementing these solutions, we should consider these disadvantages properly. The combination of identifying attacks, addressing vulnerabilities at various layers, and using countermeasures forms a strong defense to protect the security and integrity of blockchain systems.

\section{Quantum attacks and mitigation on Blockchain}

Blockchain technologies heavily depend on cryptographic protocols for many fundamental processes, such as transaction mechanisms and proving ownership. The transaction process includes moving tokens and data between participants, which requires a digital signature to confirm their ownership of the keys used to create it. Various signature schemes, such as ElGamal, RSA, and Schnorr. The Elliptic Curve Digital Signature Algorithm (ECDSA) \cite{johnson2001elliptic} is a signature scheme that hinges on the challenge of solving the discrete logarithm problem within elliptic curves. Blockchain technology is based on consensus algorithms, which work on solving complex problems. The trustless aspect of blockchains comes from these consensus methods and the cryptography supporting them. Conventional public-key cryptosystems depend on the computational complexity, such as addressing the discrete logarithm problem, to provide security within the blockchain. Similarly, hash functions solve a complex puzzle by finding a nonce value. The process of finding a nonce value can be likened to solving a search problem.

Quantum computing poses a significant risk to numerous cryptographic protocols currently used in blockchain. It is predicted that by 2035, there will be a quantum computer with the potential to compromise the traditional public-key cryptosystem and Elliptic Curve(EC) cryptography e.g. RSA2048 encryption, SHA256, and digital signature \cite{kearney2021vulnerability}. Quantum computers leverage quantum physical phenomena to significantly reduce the time needed to solve certain computational challenges by exploiting quantum superposition. Shor's algorithm \cite{ugwuishiwu2020overview}, a quantum algorithm, can efficiently factor large integers and solve discrete logarithms in polynomial time. Consequently, a quantum computer running at 10MHz could decrypt an RSA2048 cipher in approximately 42 minutes. In addition, Grover's search algorithm \cite{kwiat2000grover} enables the discovery of a solution within any search space of size N in O(sqrt(n)) time. Any NP-complete problem can be solved roughly twice as fast as any current classical algorithm.

\begin{figure}
\centering
\includegraphics[scale=0.65]{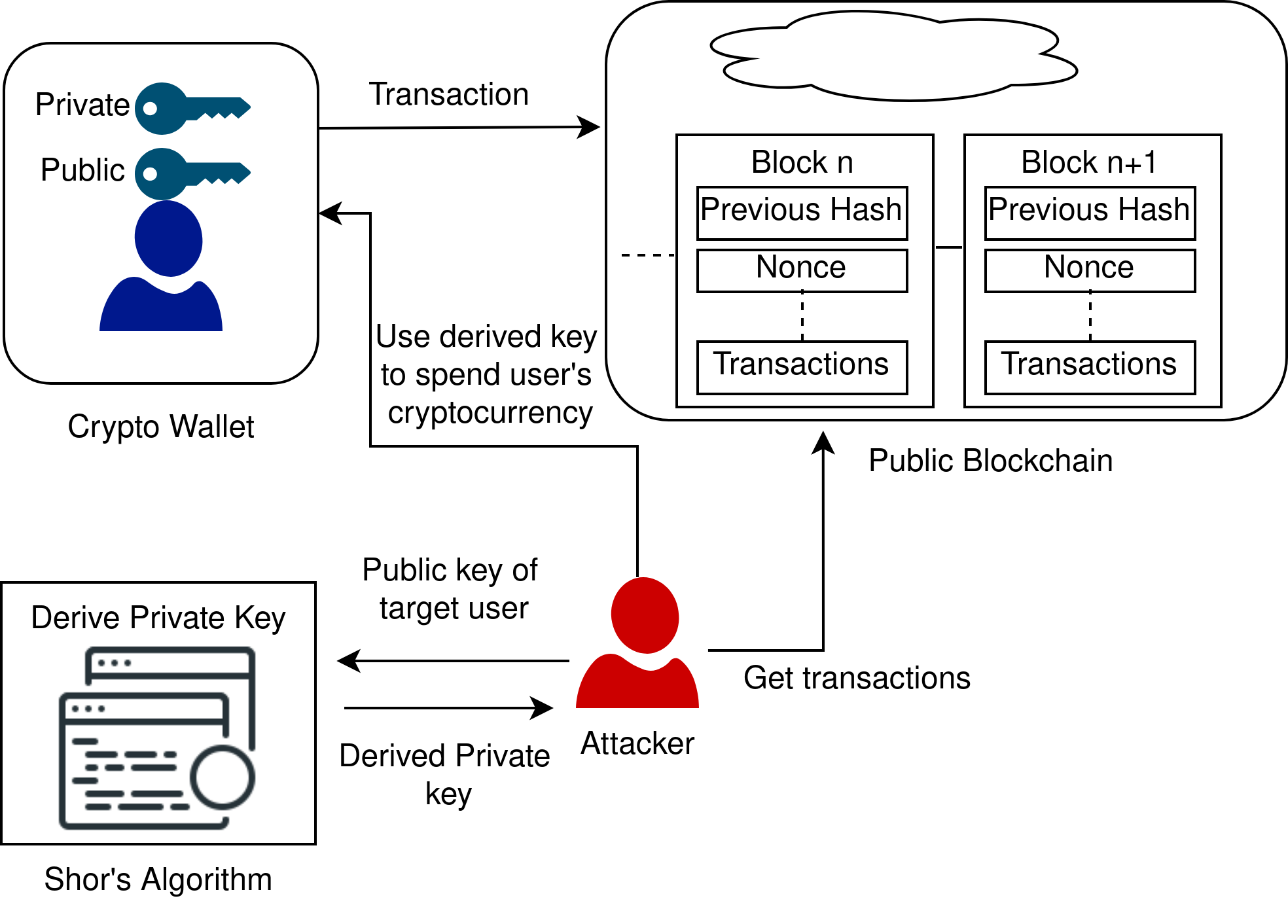}
\caption{Potential quantum attack on a blockchain}
\label{quantumattack}
\end{figure}

Possible blockchain weakness is illustrated in Figure \ref{quantumattack} by using Shor's algorithms. Even if users are actively transacting and depending on the blockchain to keep track of transactions, Shor's algorithm can help a quantum-powered attacker hijack every blockchain account. The attacker examines the public blockchain transactions and obtains the target user's public key. The attacker then uses Shor's algorithm to infer the private key. The attacker then spends the target user's cryptocurrency using the acquired private key.

On the other hand, to modify a transaction, an attacker utilizes Grover's algorithm to search for a nonce that, when applied, produces a hash that satisfies the necessary degree of difficulty. The attacker then finds a valid nonce for each of the blocks that come after the altered one to rebuild them all. The attacker then builds a sequence of fake blocks, until it is the longest chain. Other miners will respond by continuing this longer chain. It is important to note that blockchains are not effectively compromised by the current state of quantum computing capabilities since these are not strong enough. However, carrying out these attacks is expected to soon become possible.

\begin{figure}
\centering
\includegraphics[scale=0.35]{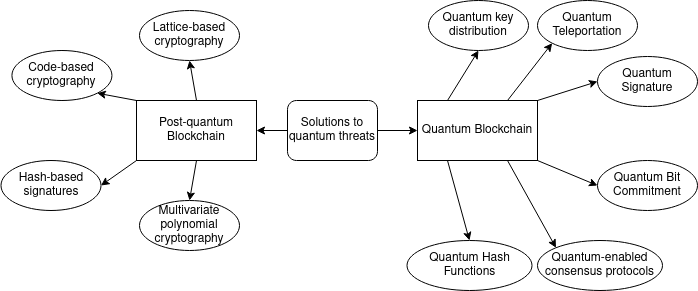}
\caption{Solutions to Quantum Threats}
\label{quantumthreatsolution}
\end{figure}

Two distinct sets of solutions have been proposed in response to the challenge posed by quantum computing, as shown in Figure \ref{quantumthreatsolution}. The first one, post-quantum blockchain, refers to cryptographic algorithms designed to be secure against attacks by quantum computers. It aims to replace or enhance existing cryptographic algorithms, such as RSA and ECDSA, which are vulnerable to attacks by quantum computers. Some examples of post-quantum cryptographic techniques are lattice-based cryptography \cite{li2018new}, code-based cryptography \cite{balamurugan2021post}, hash-based signatures \cite{butin2017hash}, and multivariate polynomial cryptography \cite{paul2022tensor}, which are being standardized to provide long-term security. The interested reader can refer to the following articles: \cite{bavdekar2023post}, \cite{xu2023overview}, and \cite{biasse2023quantum} to expand their understanding of post-quantum cryptography.

The second one, quantum blockchain, explores integrating quantum computing technologies with blockchain. It aims to leverage quantum computers' computational power and capabilities for blockchain-related tasks, such as optimization problems, hashing, and cryptography. The ongoing research is to transform the framework of traditional blockchains by researching the implementation of quantum computers and quantum networks. The quantum blockchains may solely depend on the quantum hardware or may use a hybrid architecture that integrates the components of both classical and quantum computers. They can be used to improve consensus algorithms, speed up cryptographic operations, and solve complex problems more efficiently. The research on developing quantum computers and quantum networks is ongoing. However, some protocols and algorithms that utilize classical and quantum communication have been developed. For instance, quantum key distribution (QKD) \cite{alleaume2014using} is developed for the hybrid networks for exchanging the key securely. Other quantum concepts and techniques, such as quantum teleportation, quantum signature, quantum-enabled consensus protocol, quantum hash functions, and quantum bit commitment, can serve as the foundation for building blockchains \cite{yang2023survey}. 

Post-quantum cryptography can provide security for traditional blockchains. Still, it is suggested that the complete realization of decentralized networks can only be achieved by integrating quantum blockchains into a quantum internet. The authors in  \cite{yang2023survey} compare these two solutions and mention the challenges that must be tackled in these domains. We identify the potential quantum attacks on the currently running blockchains and the available mitigation techniques.

\textbf{\textit{Mining Centralization: }} The advent of quantum computers poses a potential threat to the decentralization of blockchain networks. Many blockchains, such as Bitcoin and Litecoin, use the Proof of Work(PoW) consensus mechanism to validate transactions. This consensus mechanism involves miners to perform a computationally hard problem of calculating a SHA-256 hash value with certain conditions. This is a time-consuming process with conventional computational hardware and requires considerable computational power. This consensus mechanism is vulnerable to quantum attacks. An attacker can use Grover's search algorithm to compute this task much faster than other miners using classical machines. This would enable the attacker to perform a 51\% attack at ease. This could also lead to a significant increase in chain reorganizations and double-spending attacks. Miners or attackers with access to quantum computing could manipulate the blockchain's history, causing disruptions and leading to trust issues. With the advances in conventional mining techniques, the feasibility of performing such an attack using Grover's algorithm is low. However, the amplification of its advantage over current mining poses a serious threat.

\textbf{Mitigation}: Several researchers have suggested shifting to quantum computing-based blockchains, or quantum blockchains, to future-proof the technology. In \cite{jogenfors2019quantum} and \cite{ikeda2019qbitcoin}, the authors have proposed to migrate Bitcoin itself to quantum computers and accelerate the consensus mechanism using Grover's algorithm. Dolev et al. \cite{dolev2020sodsbc} propose an asynchronous consensus model, SodsBC, utilizing concurrent processing to attain quantum safety. Since the algorithm does not provide an exponential advantage for PoW consensus, faster developments in current mining techniques can be easily used to further the date when 51\% attacks will be feasible on PoW-based blockchains. Despite large post-quantum projects on blockchain, there has been a lack of a large initiative that can accelerate the growth of this technology. 

\textbf{\textit{Key Reuse Vulnerabilities:}} Cryptocurrency transactions are currently encrypted using digital signature algorithms and then broadcast over the network. Bitcoin, for example, uses ECDSA, which relies on the difficulty of the Elliptic Curve Discrete Logarithmic Problem. The use of quantum computers to solve the problem reduces the time complexity from exponential($O(2^n)$) to polynomial((O($n^3$)). Once the public key is known, it becomes much easier for the attacker to unmask the transactions by computing the private key, potentially putting the victim's funds at risk. This becomes especially risky for those transactions that have not yet been incorporated into the blockchain. With the knowledge of the private key, the attacker can impersonate the victim and sign a new transaction using this key. Aggawal et al.  \cite{aggarwal2017quantum} show how, using Shor's algorithm, a quantum computer with 485,550 qubits can solve the problem in under 30 minutes. Moreover, all the transactions that users store on their hot wallets become vulnerable to de-anonymity threats from an attacker, revealing sensitive information like wallet addresses and transaction details, jeopardizing the victim's privacy. 

\textbf{Mitigation}: While quantum computers using Shor's algorithm might not be an immediate threat to cryptocurrencies, it is important to understand the mitigation steps that we can take in the future to prevent such attacks in the future. Regular rotation of public and private keys for wallets is important to limit the blast radius of such an attack. If an attacker generates a private key, it might become outdated by the time they can execute their attack. Transitioning to post-quantum cryptographic solutions, which are more resilient against quantum attacks, is imminent. As mentioned before, several candidates, such as lattice-based cryptosystems, code-based cryptosystems, and multivariate polynomial-based cryptosystems, have been proposed in the past, which are much more robust against Shor's algorithm. Using multi-signature wallets, which require multiple private keys to sign a transaction, also adds a new layer of security against such quantum threats. 

\begin{figure*}
\centering
\includegraphics[width=1\linewidth]{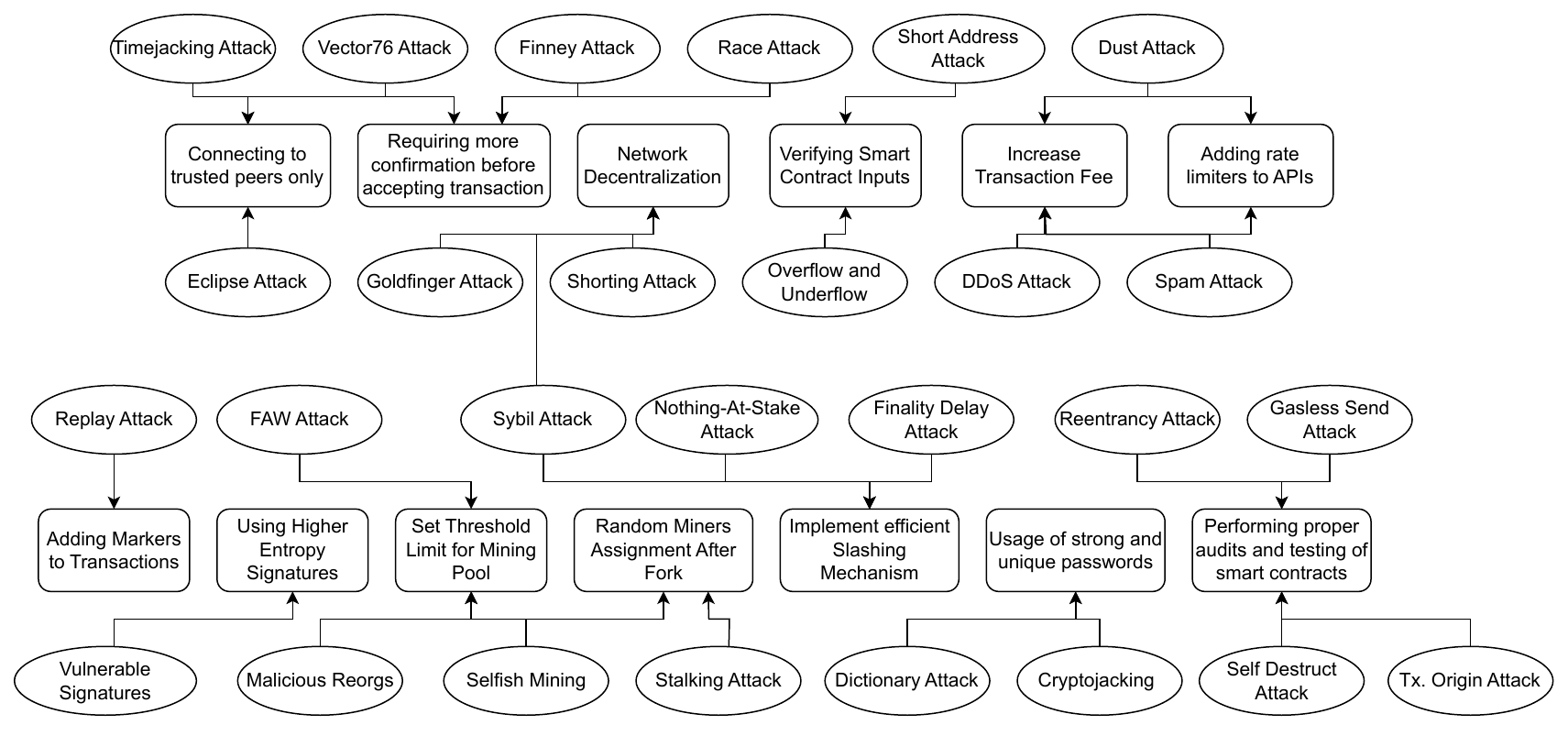}
\caption{Countermeasures against common attacks}
\label{countermeasures}
\end{figure*}

\textbf{\textit{Smart Contract Vulnerabilities:}} Smart contracts are crucial for expanding blockchain applications beyond simple P2P transactions. These contracts have the terms of the agreement written inside them and automatically execute when the mentioned conditions are met. However, some smart contracts that rely on traditional cryptographic techniques are vulnerable to quantum attacks. They use digital signatures to verify the authenticity of the transaction. Digital signatures are based on cryptographic algorithms, like ECDSA. As mentioned previously, these classical algorithms can be solved easily by quantum computers using Shor's algorithm. An attacker, using such quantum computers, can derive the private key from the public key and impersonate the user in smart contracts. If such cryptographic algorithms protect a smart contract's conditions and outcomes, an attacker can manipulate the smart contract to execute erroneously, modify contract parameters, or halt the contract's intended operation.

\textbf{Mitigation}: To mitigate these vulnerabilities, developers and blockchain platforms should consider adopting quantum-resistant cryptographic methods. Quantum-resistant cryptography aims to provide security against quantum attacks by using algorithms that are secure against powerful quantum computers. Dolev et al. \cite{dolev2021sodsbc} propose a new post-quantum smart contract system, SodsMPC, to achieve the privacy of the contract's business logic by using multi-party computation protocols. It also ensures the accuracy of the smart contract's execution while maintaining its data privacy. In  \cite{cai2019blockchain}, the authors present an innovative architecture built on smart contracts to defend against quantum attacks. Their approach relies on quantum blind signatures, which offer versatile applications for both single and multiple signatures. All the mitigation steps mentioned in the previous attack vector, Key Reuse Vulnerability, apply here as well.

\section{Future Research}
As newer technologies will emerge, threats to the existing systems will get more aggressive. Understanding and mitigating these emerging attack vectors becomes paramount. Moreover, the potential of quantum computing necessitates proactive research into post-quantum cryptography solutions. The interplay of blockchain networks demands a thorough examination of interoperability challenges and the development of secure cross-chain protocols. 

\textbf{Application Layer:} The applications working as the end product of blockchain, which its users interact with, have several shortcomings. We explored the Race Attack, Finney Attack, and Vector76 Attack, which target this layer. The proposed solution of waiting for transaction confirmation from various nodes can be useful. However, it can degrade the user experience by increasing the turnaround time, especially in slower blockchains like Ethereum and Bitcoin. Further research into making faster transaction finalization while keeping the blockchain secure needs to be done. Incidents such as the FTX collapse  \cite{jalan2023systemic} warrant the entrance of regulatory authorities in this space to enhance the trust among the public.

\textbf{Contract Layer:} Smart contracts written for adding additional functionalities to the blockchain’s transactional feature are at the complete discretion of the smart contract creator. The various attacks we discuss in this paper, like Reentrancy Attack, Gasless Send, Tx. Origin Attacks can all be countered by writing additional checks into the smart contract. However, significant research into adding such checks without increasing the gas cost and resource consumption is required. Unless we address these gaps, wider adoption of blockchain will be challenging. Maintaining security in smart contracts while allowing inter-blockchain functionality will pose newer challenges, requiring additional scrutiny.

\textbf{Consensus Layer:} The consensus mechanism in blockchain forms the crux of its decentralization principle. Any attack on this layer undermines this crucial principle. We discuss several attacks like Shorting Attack, Goldfinger Attack, FAW Attack, etc. Blockchain centralization and forking vulnerabilities need to be addressed, especially with the advent of newer blockchains having fewer users. Some blockchains, like EOS  \cite{xu2018eos}, use consensus mechanisms with lower block finality time. Further scrutiny of these novel consensus models is needed to ensure it does not introduce newer vulnerabilities. Regulatory authorities, which prevent the centralization of blockchain, though necessary, need to be kept in check to sustain the privacy and innovation in this domain. 

\textbf{Network Layer:} This layer dictates how the nodes in a blockchain interact with each other. The main threat to this layer is attackers creating malicious nodes and flooding the network with transactions, which leads to various attacks like DDoS Attacks, Sybil Attacks, Timejacking Attacks, etc. Research is required to improve peer discovery protocols to protect honest nodes from connecting to malicious ones, taking the burden of verifying connections away from the users. Newer cryptographic techniques, such as zero-knowledge proofs, must also be explored to secure the privacy of users’ data flowing over the network. Inter-blockchain connection and its secure operability should also be considered, as this would be paramount in scaling blockchain solutions to everyday problems and wider adoption.

\textbf{Data Layer:} Apart from the decentralization principle of blockchain, its privacy-preserving nature attracts a lot of people. Improper credentials or mishandling them might attract various attacks like Replay Attacks, Cryptojacking, Dictionary Attacks, etc. Research into introducing interoperability standards to ensure proper data handling with cross-chain integration is required. Research into enforcing sufficiently secure passwords would help protect users from attackers trying to pose as them. For storing transaction data, many blockchains use off-chain solutions, like IPFS (Interplanetary File Storage)  \cite{benet2014ipfs}. Several threats can emerge from the interaction between the blockchain and the database, requiring significant research. 

Figure 11 shows the various mitigation techniques that can affect the listed attacks. However, we need to use them with proper cognizance. All the mitigation steps suggested in this paper must be thoroughly examined and their ramifications considered before implementing them on a large scale. The disadvantages of all the mitigation steps are delineated in Table 2 in this paper. More research is required to address the disadvantages mentioned in this paper. Scalable solutions that sustain performance while preserving security are critical, along with developing educational resources to instill security best practices. 

The development of quantum computers makes it important for us to evaluate its impact on blockchain technology \cite{chen2021construction}. Blockchain might reach a large scale when quantum computers become widely available. Quantum computing will be able to break the current cryptographic and consensus mechanisms used by classical blockchains. Quantum computing would percolate every layer of blockchains, making it resistant to quantum computers. To enhance blockchain security in the future, a hybrid solution with a combination of quantum and classical computing will be required. Significant research in these areas is paramount.

Due to the fledgling nature of blockchain, many attacks have not yet surfaced. It is essential to weed out such attacks through extensive research before they occur in the real world. To make a truly resilient large-scale trustless ecosystem of transactions, all the layers of blockchain have to be thoroughly examined and vulnerabilities fixed at their core instead of relying on mitigation steps alone.

\section{Conclusion}

In this comprehensive review article, we conduct a thorough analysis of blockchain security, examining historical breaches and potential future threats. We classify the attacks based on the blockchain layered architecture, unveiling the inherent vulnerabilities within each. We also discuss a set of mitigation strategies for the attacks at each layer. This novel classification framework not only unravels the origins and dimensions of these threats but also equips us with a strategic defense.

In this paper, we explore the various aspects of blockchain security challenges. Through our classification, we pinpoint the specific weak points that various attacks exploit in each layer. The insights derived from this survey can be used by researchers to focus their efforts on addressing these vulnerabilities plaguing the different layers of blockchain. We found that despite the robustness of the core principles of blockchain, its implementation has a lot of weaknesses. As the influence of blockchain technology continues to grow, addressing these security concerns becomes increasingly critical. Recent high-profile attacks serve as stark reminders of this fact. To mitigate these challenges, a prudent approach to the discussed mitigation strategies is essential, considering their disadvantages, as mentioned in this paper. This would ultimately lead to the eradication of the attacks mentioned in this paper. While we discussed many attacks and vulnerabilities in this survey, the most pressing issue is blockchain centralization or 51\% attacks, considering their blast radius and the fact that we do not have a foolproof solution to it yet. Newer blockchains with limited users need to be especially vigilant in their implementation since such an attack might destroy the entire network.

One of the limitations of this study is that it does not provide qualitative results or in-depth security analysis to measure the performance of existing countermeasures techniques protecting the blockchain from specific attacks. Further research into measuring the performance of the various mitigation strategies suggested in this paper will help blockchain developers create more resilient solutions. The evolving landscape of technology and threats means that new vulnerabilities may emerge. The mitigation steps mentioned in this paper might not prevent attackers from exploiting such new vulnerabilities. 

Our exploration into the implications of quantum computing on blockchain highlights an impending paradigm shift in the field. Although quantum computing is not an immediate threat to blockchain technology, research endeavors must focus on quantum and post-quantum blockchain technologies to safeguard its future. Our work lays the foundation for informed strategies in fortifying these systems, ultimately enhancing user trust and facilitating its responsible integration across various sectors and applications.



\end{document}